\documentclass[fleqn,usenatbib]{mnras}

\usepackage{newtxtext,newtxmath}

\usepackage[T1]{fontenc}

\DeclareRobustCommand{\VAN}[3]{#2}
\let\VANthebibliography\thebibliography
\def\thebibliography{\DeclareRobustCommand{\VAN}[3]{##3}\VANthebibliography}

\usepackage[utf8]{inputenc}
\usepackage{graphicx}	
\usepackage{amsmath}	

\usepackage{amssymb}	
\usepackage{xcolor}
\usepackage{xspace}     
\usepackage{multirow}  
\usepackage{longtable} 

\newcommand{\angstrom}{\text{\normalfont\AA}\xspace} 
\newcommand{\degree}{\ensuremath{^{\circ}}\xspace} 
\newcommand{\umt}{\text{\normalfont\textmu m}\xspace} 

\graphicspath{{figures/}}

\title[Planets in photoevaporating discs: FEL diagnostics]{The interplay between forming planets and photo-evaporating discs I: Forbidden line diagnostics}

\author[Weber et al.]{
Michael L. Weber$^{1,2}$\thanks{E-mail: mweber@usm.lmu.de (MLW)},
Barbara Ercolano$^{1,2}$,
Giovanni Picogna$^{1}$,
Christian Rab$^{1,3}$
\\
$^{1}$University Observatory, Faculty of Physics, Ludwig-Maximilians-Universität München, Scheinerstr. 1, 81679 Munich, Germany\\
$^{2}$Excellence Cluster 'Origins', Boltzmannstr. 2, 85748 Garching, Germany\\
$^{3}$Max-Planck-Institut für extraterrestrische Physik, Giessenbachstr. 1, 85748 Garching, Germany
}

\date{Accepted XXX. Received YYY; in original form ZZZ}

\pubyear{2022}

\begin{document}
\label{firstpage}
\pagerange{\pageref{firstpage}--\pageref{lastpage}}
\maketitle

\begin{abstract}
Disc winds and planet formation are considered to be two of the most important mechanisms that drive the evolution and dispersal of protoplanetary discs and in turn define the environment in which planets form and evolve. While both have been studied extensively in the past, we combine them into one model by performing three-dimensional radiation-hydrodynamic simulations of giant planet hosting discs that are undergoing X-ray photo-evaporation, with the goal to analyse the interactions between both mechanisms. In order to study the effect on observational diagnostics, we produce synthetic observations of commonly used wind-tracing forbidden emission lines with detailed radiative transfer and photo-ionisation calculations.
We find that a sufficiently massive giant planet carves a gap in the gas disc that is deep enough to affect the structure and kinematics of the pressure-driven photo-evaporative wind significantly. This effect can be strong enough to be visible in the synthetic high-resolution observations of some of our wind diagnostic lines, such as the [OI]~6300~\angstrom or [SII]~6730~\angstrom lines. When the disc is observed at inclinations around 40\degree and higher, the spectral line profiles may exhibit a peak in the redshifted part of the spectrum, which cannot easily be explained by simple wind models alone.
Moreover, massive planets can induce asymmetric substructures within the disc and the photo-evaporative wind, giving rise to temporal variations of the line profiles that can be strong enough to be observable on timescales of less than a quarter of the planet's orbital period.

\end{abstract}




\section{Introduction}\label{sec:introduction}
As the birth sites of planets, protoplanetary discs (PPDs) provide not only the material out of which planets form, but also the environmental conditions in which they form and then evolve. Observations suggest that PPDs typically disperse over timescales of only a few million years \citep[e.g.][]{Haisch2001,Fedele2010,Ribas2014} and the observed disc fractions at different wavelengths hint at an inside-out dispersal of discs \citep[see also][]{Koepferl2013, Ercolano2015}. Understanding the processes that drive this evolution and dispersal is essential to understand the formation and evolution of the many diverse planets and planetary systems that have been discovered in recent years.

While several mechanisms have been proposed to be contributing to the evolution of PPDs, it remains difficult to constrain their relative importance. Angular momentum transport through turbulence, induced by magnetorotational instability (MRI) \citep{Balbus1991}, has long been believed to be the dominant process for driving viscous accretion, but viscous accretion by itself cannot explain the inside-out dispersal of discs. Moreover, advancements in theoretical models have brought the effectiveness of MRI into question, as it became clear that MRI-driven turbulence is greatly reduced at several disc locations when non-ideal magnetohydrodynamical (MHD) effects are considered \citep{Bai2015, Simon2018}. It is very likely that the main mechanism responsible for driving accretion varies with the location in the disc and possibly its evolutionary stage. Growing evidence suggests that magnetic disc winds play an important role by extracting angular momentum from the disc \citep{Gressel2015a,Bai2017,Bethune2017,Wang2019,Gressel2020a}.
Thermal disc winds, while unable to drive accretion, contribute directly to disc dispersal by carrying away material from the disc surface with mass-loss rates on the order of $10^{-8}$~M$_\odot$/yr that can lead to fast inside-out dispersal of a disc \citep[see][for recent reviews]{Ercolano2017a, Pascucci2022}. With these considerations, both thermal and magnetic disc winds appear to play one of the most important roles in disc evolution.

Another important mechanism is the interaction between the disc and massive planets within. A sufficiently massive planet that is embedded in a PPD will exert enough gravitational torque on the surrounding material that a gap is opened along the planet's orbit \citep{Papaloizou1984} and spirals are generated in the disc \citep{Ogilvie2002}, perturbing its gas and dust structure. Many spatially resolved observations of PPDs show detailed substructures such as gaps, rings and spirals \citep{ALMAPartnership2015, Andrews2018} and often planet-disc interactions are proposed as their most likely origin. In fact, some of the substructures have recently been associated to kinematic detections of planets \citep[e.g. review by][and references therein]{Pinte2022}. Planet-disc interactions are also the driver for planet migration and therefore crucial to understand the evolution of newly formed planets and the architecture of planetary systems. 
Numerical models show that planet migration can strongly be affected by disc dispersal via photo-evaporative winds \citep{Alexander2009, Ercolano2015a, Jennings2018b, Monsch2021b} and in an observational study of the semi-major axis $a$ of exoplanets and the X-ray luminosity L$_X$ of their host stars \citet{Monsch2019} found a void in the $L_X-a$ plane that may hint towards the parking of giant planets close to where a wind launched by X-ray photo-evaporation would open a gap. \citet{Alexander2009} and \citet{Rosotti2013} have found numerically that a planetary gap may reduce the mass inflow into the inner disc and in consequence accelerate dispersal of the inner disc by a photo-evaporative wind. The effect is referred to as planet-induced photo-evaporation (PIPE). This demonstrates the importance that the interplay between disc winds and planet-disc interactions has for the evolution of PPDs. 

While in the recent years spatial observations of discs and the substructures within them have become increasingly more detailed, direct observations of disc winds remain challenging. One of the most important observational tools for studying the structure and kinematics of disc winds are forbidden emission lines (FEL). These are usually lines from neutral or lowly ionized elements, such as the [OI]~6300~\AA \, or [NeII]~12.8~\umt lines, but also molecular low-velocity outflows from the inner disc regions have been detected in CO emission \citep[e.g.][]{Bast2011,Pontoppidan2011} or H$_2$ emission \citep{Gangi2020}. The lines often show complex spectral profiles that are blueshifted with respect to the stellar velocity, indicating an outflow that is approaching the observer. The profiles can provide insight into the structure of the outflow and its origin. To this end, they are often decomposed into one or multiple Gaussian components and classified according to the centroid velocity and width of the components. It should be noted that a single component does not necessarily trace the emission from a specific physically closed volume of the wind \citep[see also][]{Weber2020}, but they are nevertheless useful to build statistics and study kinematic links. High-velocity components (HVC) with blueshifts $\gtrapprox$~30km/s are usually attributed to fast-flowing jets, while less blueshifted low-velocity components (LVC) are believed to have their origin in slower winds \citep{Hamann1994, Hartigan1995}. The LVCs can be further decomposed into broad components (BC) with FWHM $\gtrapprox$~40km/s and narrower components (NC) \citep{Rigliaco2013,Simon2016a,McGinnis2018,Fang2018,Banzatti2018}. Under the assumption of Keplerian-broadening, BCs are often believed to trace winds within the inner 0.5~au of the disc and thus to have their origin in a magnetic wind \citep{Simon2016a}, while narrow components (NC) are attributed to outflows at larger radii and are consistent with both, MHD winds \citep{Banzatti2018} and theoretical models of PE winds \citep{Ercolano2010,Weber2020,Ballabio2020}. \citet{Pascucci2020} suggest an evolution in time, where inner MHD winds are dominating at early stages of disc evolution and outer winds could become more important at later stages. On short timescales ($\sim$1~decade) the overall structure of the LVC remains mostly stable within the limits of the available spectral resolution \citep{Simon2016a}, but for a few objects there exist multiple spectra with comparable resolution ($\sim$6~km/s) that were observed a few years apart and show small variations in the LVC \citep[e.g. UX Tau in][]{Simon2016a, Fang2018}.

To understand and interpret the observed line profiles, synthetic observations of theoretical models that can be compared to real observations have proven useful. While it is still numerically challenging to calculate synthetic line profiles for MHD wind models, extensive work in that direction has been done on hydrodynamical models of PE winds. \citet{Font2004} modelled spectral profiles and fluxes of various optical forbidden lines emitted in an EUV-driven photo-evaporative wind and found good agreement with observational data for lines of ionized species, but their models could not account for the luminosities of neutral oxygen lines. \citet{Alexander2008} extended this work by modelling the mid-infrared [NeII]~12.8~\umt line and found encouraging results, too. \citet{Ercolano2010} used the X-ray+EUV photo-evaporation models by \citet{Owen2010} to predict many optical and mid-infrared lines and found good agreement with the observed line luminosities, including that of the [OI]~6300~\angstrom neutral oxygen line and subsequently \citet{Ercolano2016} showed that the observed correlation between line intensity and accretion luminosity can be well reproduced by adding an additional accretion luminosity component to the irradiating spectrum. Later \citet{Picogna2019} introduced a new generation of X-ray photo-evaporation models that improved on the model by \citet{Owen2010}. With  those, \citet{Picogna2019} and \citet{Weber2020} showed that many of the observed NC can be reproduced, as well as their correlations with the observed BC although the BC is not present in a PE wind and is likely to have a different origin.

This paper is the first of a series in which we aim to connect two important drivers of disc evolution -- disc winds and planet-disc interactions -- and study their interplay by means of three-dimensional radiation-hydrodynamic simulations of X-ray photo-evaporating protoplanetary discs that host a giant planet. In this paper, we present synthetic forbidden emission line profiles calculated with these models and study the effects that the presence of the giant planet has on the wind and on the forbidden emission lines that are emitted in it.

\section{Methods}\label{sec:methods}
\subsection{Hydrodynamical models} \label{sec:met:models}

\subsubsection{2D r-$\theta$ model of a disc with X-ray photo-evaporation} \label{sec:met:models:2d}

As a first step, we have rerun the two-dimensional model of an X-ray photo-evaporating primordial protoplanetary disc with a 0.7 M$_\odot$ host star by \citet{Picogna2019}. This model uses a temperature parametrisation that is derived from detailed radiative-transfer calculations performed with the gas and dust radiative transfer code MOCASSIN \citep{Ercolano2003, Ercolano2005a, Ercolano2008}. The radiative transfer calculations used the X-ray + EUV spectrum presented by \citet{Ercolano2008a, Ercolano2009} with a luminosity of $L_X = 2\cdot10^{30}$ erg/s and solar abundances from \citet{Asplund2005}, depleted according to \citet{Savage1996}. The resulting parametrisation depends on the local ionisation parameter $\xi = L_X / nr^2$ and the gas column number density $N_H$ along the line of sight towards the star, both of which are easily accessible in the hydrodynamical simulation, allowing  for very efficient temperature updates at each hydrodynamical time step, while retaining good accuracy to the full radiative-transfer calculations \citep[fig. 1]{Picogna2019}. The parametrisation is applied up to a gas column number density of 2.5$\cdot10^{22}$ cm$^{-2}$, beyond which thermal coupling between gas and dust is assumed and the temperature is set to the dust temperatures from the models of \citet{DAlessio2001}. The initial density and temperature distribution is taken from the hydrostatic equilibrium models of \citet{Ercolano2008a, Ercolano2009}. To reduce computational cost, the disc is assumed to be symmetric with respect to the disc midplane. We reran this model using the PLUTO code \citep{Mignone2007} for 500 orbits at 5.2~au, after which a steady state has been reached. It serves as a starting point for our 3D models with a planet, as well as a reference model for a disc with no planet. The parameters for this and all other models are listed in table \ref{tab:model-params}. The details of the numerical grids used in the hydrodynamical simulations are listed in table \ref{tab:grids}. 

\begin{table}
\centering
\caption{Model parameters for the hydrodynamical models. We name the 2D reference model with no planet REF, the other two models that include a planet are named MJ1 and MJ5, according to the planet's mass.}
\label{tab:model-params}
\begin{tabular}{lll}
    Parameter                       & Value                     & Model-ID \\
    \hline
    \hline
                                    & 0                         & REF \\
    M$_\mathrm{planet}$             & 1~M$_\mathrm{J}$        & MJ1 \\
                                    & 5~M$_\mathrm{J}$        & MJ5 \\
    \hline
    M$_*$                           & 0.7~M$_\odot$             & \\
    M$_\mathrm{disc}$               & $\approx$ 0.1~M$_*$       & \\
    $\alpha$ (viscosity parameter)  & $10^{-3}$                 & \\
    $\gamma$ (adiabatic index)      & 1.4                       & \\
    $\mu$ (mean molecular weight)   & 1.37125                   & \\
    L$_\mathrm{X}$ [erg/s]          & $2 \cdot 10^{30}$ erg/s   &
\end{tabular}
\end{table}

\begin{table*}
\centering
\caption{Grid composition for the hydrodynamical simulations. The intervals in the radial dimension are given in code units (1~c.u. = 5.2~au).}
\label{tab:grids}
\begin{tabular}{l|lll|lll|lll}
    \multicolumn{1}{c}{} & \multicolumn{3}{c}{Step 1: 2D polar} & \multicolumn{3}{c}{Step 2: 3D spherical} & \multicolumn{3}{c}{Step 3: 3D spherical, extended}  \\
    Dim. & Interval & N Points & Type & Interval & N Points & Type & Interval & N Points & Type \\
    \hline
    \hline
    \multirow{3}{*}{r} & & &              & [0.28, 0.846] & 60 & uniform        & [0.077, 0.846] & 80 & uniform \\
    & 412 & [0.06, 115] & logarithmic     & [0.846, 1.154] & 40 & uniform       & [0.846, 1.154] & 40 & uniform \\
    & & &                                 & [1.154, 2.5] & 60 & logarithmic     & [1.154, 2.5] & 60 & logarithmic \\
    \hline
    
    \multirow{2}{*}{$\theta$} & \multirow{2}{*}{160} & \multirow{2}{*}{[0.005, $\frac{\pi}{2}$]}  & \multirow{2}{*}{uniform}     & [0.3, 1.417] & 68 & uniform                & [0.088, 1.417] & 82 & uniform \\
    & & &                                                                        & [1.417, $\frac{\pi}{2}$] & 20 & uniform     & [1.417, $\frac{\pi}{2}$] & 20 & uniform \\
    \hline
    
    & & &           & [0, 0.154] & 20 & uniform         & [0, 0.154] & 20 & uniform \\
    $\Phi$ & & &    & [0.154, 6.130] & 355 & uniform    & [0.154, 6.130] & 355 & uniform  \\
    & & &           & [6.130, 2$\pi$] & 20 & uniform    & [6.130, 2$\pi$] & 20 & uniform \\
    \hline
\end{tabular}
\end{table*}

\subsubsection{3D model of a disc with X-ray photo-evaporation and a giant planet} \label{sec:met:models:3d}

In a second step, we extended the 2D model into a three-dimensional spherical grid, assuming azimuthal symmetry around the polar axis. After verifying that the 3D-model yields exactly the same results as the 2D-model, we introduced an additional gravitational potential at ($R_p$, $\theta_p$, $\phi_p$) = (1, $\frac{\pi}{2}$, 0) which we adopted from \citet{Klahr2006}:
\begin{align}
    \Phi_\mathrm{P} =   
    \begin{cases}
        -\frac{G~M_p}{d} \left[ \left( \frac{d}{r_{sm}}\right)^4 - 2 \left( \frac{d}{r_{sm}}\right)^3 + 2 \frac{d}{r_{sm}} \right],& \text{if } d\leq r_{sm}\\
        -\frac{G~M_p}{d},              & \text{if } d > r_{sm}
    \end{cases}.
\end{align}
This models the gravitational potential of a planet with mass $M_p$, that is smoothed with a cubic polynomial if the distance $d$ from the planet is less than the smoothing length $r_{sm}$. Following \citet{Kley2009} we chose a smoothing length of 0.5 R$_\mathrm{Hill}$. Converted from code units, $R_p = 1$ corresponds to a planet with an orbital distance of 5.2 au, comparable to Jupiter in the solar system. 
To prevent strong disturbances of the disc, we used a sinusoidal function to slowly increase the planet mass from 0 to its final value $M_p$ over 20 planetary orbits.
In order to more accurately capture the planet-disc interactions, we used a nested, rotating grid, to achieve a higher resolution of 0.1 R$_\mathrm{Hill}$ within a region extending $\approx2$ R$_\mathrm{Hill}$ around the planet's location (see table \ref{tab:grids}). The planet's orbit was held fixed. At the radial boundaries we employed reflective conditions, in the polar direction we used a mirror condition at the disc midplane and an open boundary in the disc wind. In the azimuthal direction periodic boundary conditions were applied. We note that the choice of a reflective boundary at the midplane greatly reduces computational cost, but has the disadvantage that the planet is placed directly on the domain boundary. In theory, this could lead to numerical problems that could affect the density structure close to the planet or the propagation of the planet-induced spiral waves. However, it is unlikely that this has strong effects with the massive planets considered in this work and we do not expect it to have a significant effect on the depth and width of the gap or the disc surface, which is the most relevant aspect when studying the interaction between the planet-induced substructures and the wind. To prevent reflections of the spiral waves, we implemented wave-killing damping regions in the radial intervals [0.280, 0.376] and [2.308, 2.5] in code units, similar to those of \citet{DeVal-Borro2006}. Inside these intervals the density and the radial and polar velocities are damped towards their respective initial values on a damping timescale of 3\% of the orbital period at the boundary.

\begin{figure}
    \centering
    \includegraphics[width=.48\textwidth]{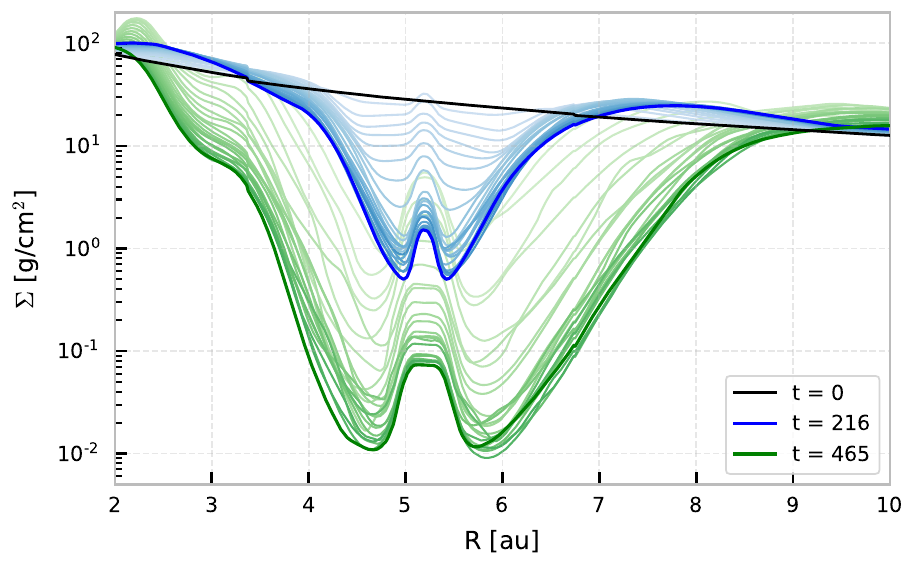}
    \caption{
        Surface density in steps of 10 orbits starting at the time of planet insertion (t = 0), which is represented by the lightest blue line. The dark blue line at t = 216 orbits represents the surface density of the MJ1 model. At the same time, it serves as the starting point for the MJ5 model (green lines), which was evolved for another 249 orbits until t = 465.
    }
    \label{fig:sigma-evol}
\end{figure}

We evolved the model until the planet-induced substructures reached a sufficiently steady state everywhere outside the planet-induced cavity, where the wind is launched. Figure \ref{fig:sigma-evol} shows the surface density evolution of the models. The model with a Jupiter-like planet (MJ1) reached a sufficiently steady state after approximately 150 orbits. We ran the simulation for 216 orbits, after which we gradually increased the planet mass over 20 orbits to 5~M$_\mathrm{J}$ for the MJ5 model. The MJ5 model was then run for another 249 orbits until t = 465. We note that there are still small variabilities in the surface density profile at the outer edge of the gap. However, these are mainly caused by hydrodynamical instabilities that lead to asymmetric features such as vortices, which are strongest close to the disc-midplane and not at the disc surface, where the wind is launched.

As a third step, we then extended the grid in the radial and polar directions by reducing the inner boundary to $R_{in} = 0.077$ (corresponding to 0.4 au) and $\theta_{in} = 0.088$ and mapping the grid from the 2D model into the newly opened space. The interval of the inner damping region was adjusted to [0.077, 0.135]. The model was then evolved for another 10 planetary orbits, which corresponds to $\approx470$ orbits at the new inner boundary. This is done to ensure that the domain includes the inner disc atmosphere which may have a small contribution to the line profiles, as well as partially screen irradiation from the central star, reducing the amount of irradiation that reaches the outer disc regions.


\subsection{Forbidden emission lines with MOCASSIN} \label{sec:met:mocassin}

We post-processed snapshots of the density and velocity grids by remapping them to a three-dimensional cartesian grid with 196 points in x- and y-direction and 86 points in the vertical z-direction, all spaced according to a power law with exponent $q = \frac{3}{4}$. Converted from code units, the remapped grid extends from -12 to 12 au in x- and y- direction and 0 to 10 au in z-Direction, however inside a radius of 0.4 au all cells are set to 0. We do not expect this region to contribute to the total line flux in a significant way, because the volume of this region is small. For our 2D model, the third dimension is not necessary, which allowed us to verify that the resolution is sufficiently high, by rerunning the 2D-calculations with a resolution that was increased by more than a factor of 6. The resulting line profiles were not affected, as the differences are not significant enough to be recognisable after the artificial degradation of the spectrum, which we employ to simulate a realistically achievable spectral resolution. 

Furthermore, all cells with a column number density $>~10^{23}$~cm$^{-2}$ are set to 0 in order to reduce computational demand. This is possible, because this condition is only met within the bound disc, where the lines that we are interested in are not emitted. 

We used the remapped cartesian grids to calculate the photo-ionisation structure and line emission with MOCASSIN, following \citet{Ercolano2010}, \citet{Ercolano2016} and \citet{Weber2020}, where we updated the atomic database to CHIANTI version 9 \citep{Dere1997,Dere2019}. The irradiating spectrum is the same as in model PE-1 of \citet{Weber2020}. It consists of a soft X-ray component with a luminosity of $2\cdot10^{30}$ erg/s and a blackbody component at a temperature of 12000~K with a bolometric luminosity L$_{\mathrm{acc}} = 2.6\cdot10^{-2}$~L$_{\odot}$, which corresponds to an ionising EUV luminosity (h$\nu$ > 13.6~eV) of $8.56\cdot10^{28}$~erg/s. This accretion component is used to model a spectrum that could be emitted by material that is accreting onto the central star. It plays an important role for the forbidden line emission along the inner edge of the wind. Since the wind is optically thick to EUV photons, it is absorbed quickly along that edge, where it dominates the ionization and heating process, but it has a negligible effect on the wind dynamics, because it is unable to penetrate the wind and to reach the wind-launching region at the disc surface \citep{Owen2012}.

\subsection{Line profile calculations} \label{sec:met:profiles}

As our set of synthetic forbidden emission lines, we chose the commonly observed [OI]~6300~\AA, [SII]~4068~\AA, [SII]~6730~\AA\, and [NeII]~12.8~\umt lines, as well as the [FeII]~17.2~\umt line and created luminosity grids using the post-processing tools that are provided with MOCASSIN. Due to the nature of Monte-Carlo simulations, the results typically contain several individual zones that are undersampled after the simulation and thus yield unrealistic values for the line luminosity. We filtered out these outliers by calculating for each zone the local mean value using the 3 nearest-neighbours in all directions and setting its value to the local median, if it exceeds the mean by 3 standard deviations. 

With the luminosity grids we were then able to calculate spectral line profiles, following the work of \citet{Ercolano2010}, \citet{Ercolano2016} and \citet{Weber2020}: 

For each volume element in the grid, we calculated the gas-velocity projected along the line of sight $v_{los}$ and the thermal rms velocity $v_{th} = \sqrt{k_B T / m}$ of the emitting element, where $m$ is the mass of the element. We then calculated the total luminosity that can be observed along the line of sight with velocity $u$ by summing the contribution of the individual zones of the 3D-grid $l(\vec{r})$:
\begin{align}
	L(u) = \int{d\vec{r} \frac{l(\vec{r})}{\sqrt{2\pi v_{th}(\vec{r})}}} exp\bigg( - \frac{[u - v_{los}]^2}{2 v_{th}(\vec{r})^2} \bigg),
\end{align}
where $\vec{r}$ is the vector pointing to the node.
This way, we created spectral line profiles by calculating the total line luminosity at a range of velocity bins spaced 0.25 km/s apart. The profiles were then convolved with a Gaussian of width $\sigma = c / R$, to simulate a finite resolving power $R$. For the optical forbidden emission lines, we chose a resolving power of $R = 50000$, which is comparable to currently available high resolution spectral observations and $R = 100000$ for profiles that are more representative of potential future observations. For the lines in the mid-infrared we used $R = 30000$ and $R = 60000$. Upcoming instruments such as ANDES on the ELT could realistically achieve a comparable resolution for the optical lines. Unfortunately, it is unlikely that the lines in the mid-infrared will be observable with such high resolution in the near future. Nevertheless, we include the profiles for completeness.

We assumed that only the disc midplane is optically thick to the lines and neglected scattering and absorption processes of the forbidden line emission which should be negligible along most lines of sight that do not cross the midplane of the disc. This approximation is rendered necessary by the limited extent of our computational domain that does not allow for an accurate calculation of the optical depth through the wind.
We created fits to our profiles using the multi-Gaussian decomposition method described in \citep{Weber2020}. This produces fits that are similar to the ones typically used for real observations. It works by adding an additional Gaussian component only if it improves the $\chi^2$-value by at least 20\%. Moreover, it includes an additional exit-condition $\chi^2 < \frac{2}{3} y^2_{max}$, where y$_{max}$ is the flux at the peak of the profile, in order to account for the fact that our synthetic profiles are much smoother and therefore generally better fitted by Gaussian components than it is the case with real spectra.

\section{Results \& Discussion}\label{sec:results}
\subsection{Wind structure and kinematics} \label{sec:res:structure}
\begin{figure*}
    \centering
    \includegraphics[width=\textwidth]{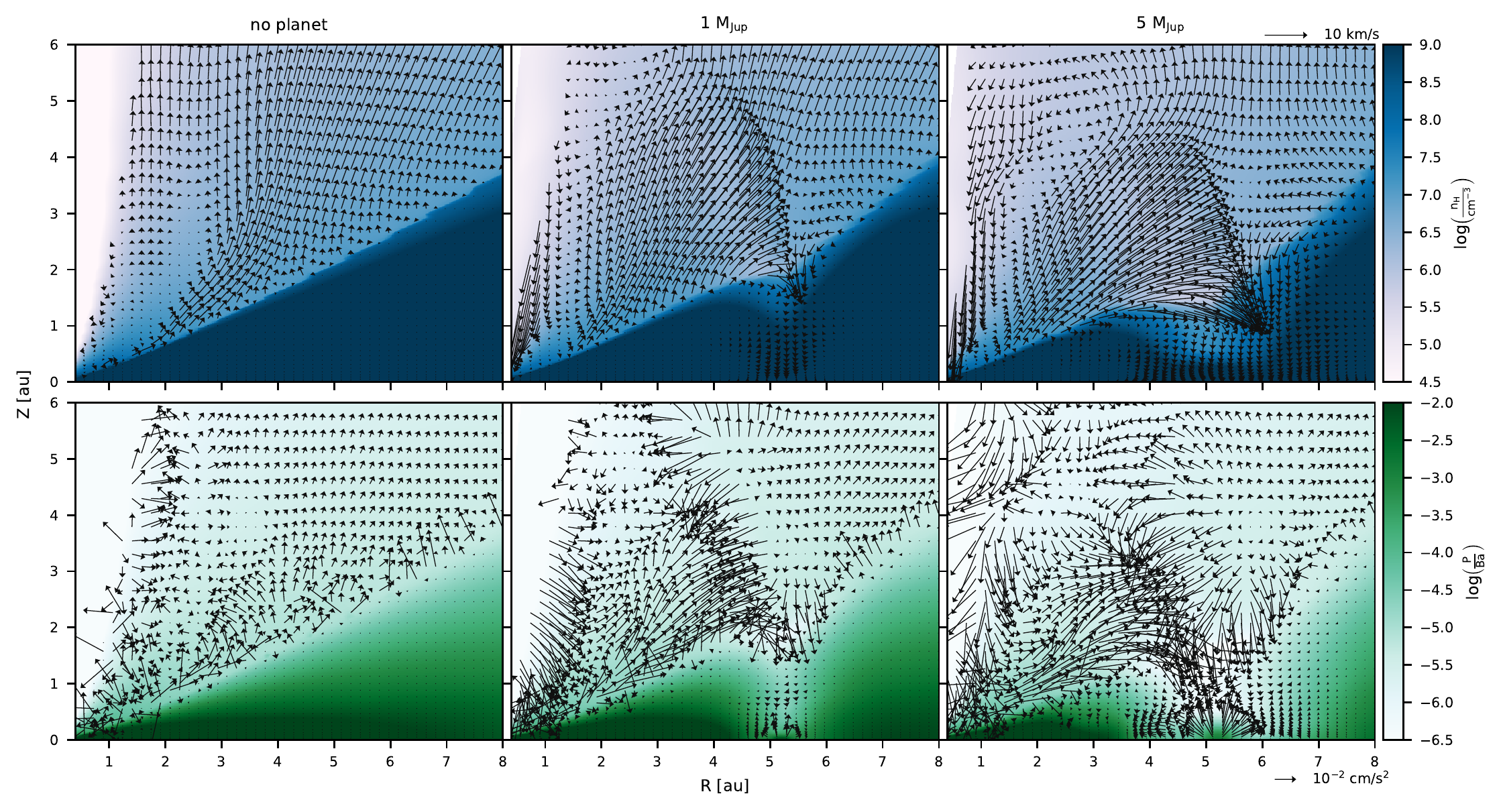}
    \caption{
        Top panels: Azimuthally averaged hydrogen number density (background color) and velocity field (black arrows).
        Bottom panels: Azimuthally averaged gas pressure (background color) and the effective acceleration as the sum of the gravitational acceleration towards the central star, the acceleration caused by the pressure-gradient force and the centrifugal acceleration.
        The velocity and acceleration fields are only plotted where the number density exceeds $2\cdot10^5$~cm$^{-3}$.
    }
    \label{fig:wind-structure}
\end{figure*}
A photo-evaporative wind is a thermal wind that is driven away from the surface of the bound protoplanetary disc by a pressure gradient. Its kinematics are therefore dependent on the density and temperature structure of the disc surface. Since massive planets interact strongly with the gas in the disc, opening up a deep gas-cavity around the entire azimuth, they could also indirectly influence a thermal disc wind.
The top panels of Figure \ref{fig:wind-structure} show the gas density structure, as well as the velocity field of the wind in our three different models. The bottom panels show the gas pressure and the effective acceleration field that we calculated by summing the three major contributing components: the gravitational acceleration towards the star, the acceleration by the pressure-gradient force and the centrifugal acceleration due to Keplerian rotation, which we have to consider in our frame of reference. The gravitational acceleration by the planet can be neglected, as in the wind regions it is much weaker than that of the star. Even in the y=0 plane at the disc surface directly above the 5~M$_\mathrm{J}$ planet, the vertical component caused by the planet reaches only $\approx 18\%$ of the stellar value. The left panels show our model without a planet, where the wind is launched consistently along the disc surface. Inside R $\approx$ 3 au, the pressure gradient is not strong enough for the wind to overcome the gravitational potential, except for a thin layer directly above the disc surface, where material can flow radially outwards, where it can later be vertically driven away from the disc. In terms of the gravitational radius $r_g$, this corresponds to a minimum launching radius of $\approx 0.1 - 0.3 r_g$, which is consistent with the critical radius found in previous studies \citep[e.g.][]{Liffman2003, Dullemond2007}.

\begin{figure}
    \centering
    \includegraphics[width=.48\textwidth]{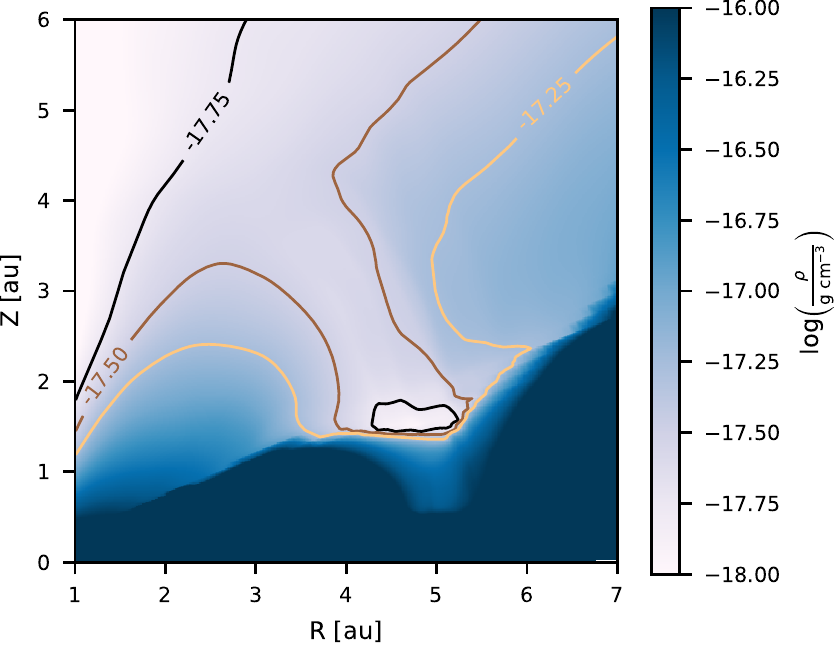}
    \caption{
        Azimuthally averaged gas density of the 5~M$_\mathrm{J}$ model. 
    }
    \label{fig:density-closeup-mj5}
\end{figure}

Comparing this to the models that include a planet-induced gap in the disc, it becomes evident that at the position of the cavity, the pressure gradient is inverted and no material is launched into the wind. This, in consequence, leads to an under-density at the location of the cavity that is perpendicular to the disc surface and reaches several au into the wind. Instead of being driven vertically farther away from the disc, material that crosses this under-density can fall back into the cavity or onto the outer edge of the gap. This effect is stronger in the model with a 5~M$_\mathrm{J}$ planet, which is able to create a much deeper cavity. In Figure~\ref{fig:density-closeup-mj5} we visualize the under-density of model MJ5. The Figure shows the azimuthally averaged gas density, where a region above the gap, spanning approximately 2 au in the radial and 4 au in the vertical direction, is apparent, within which the density is reduced by up to a factor $\approx 3$ compared to the surrounding environment.

Apart from the under-density, Figure \ref{fig:wind-structure} shows an additional effect: In the region inside R $\approx 3$~au, where in the REF model the material is mostly gravitationally bound, in the models with a planet the pressure gradient is increased towards the under-density, allowing the wind to be launched from slightly closer in but with a less steep angle. This, in turn, modifies the pressure gradients along the inner vertical edge of the wind, which is especially evident in the 
MJ5 model, where material at z$\approx$5~au reverses direction and falls back down towards the disc along the inner edge of the wind.

We note again that these effects are a consequence of the modified pressure gradient that drives the wind and are not caused by the additional gravitational potential, as discussed earlier in this section. It is not yet known whether a wind that is not dominantly driven by a pressure gradient (i.e. a magnetically driven wind), would experience similar effects on its velocity structure. Detailed numerical models would be required in order to investigate how the presence of a massive planet can affect a magnetic wind.

\subsection{Forbidden line diagnostics} \label{sec:res:line-diagnostics}
\subsubsection{Emission regions} \label{sec:res:emission-maps}
\begin{figure}
    \centering
    \includegraphics[width=.48\textwidth]{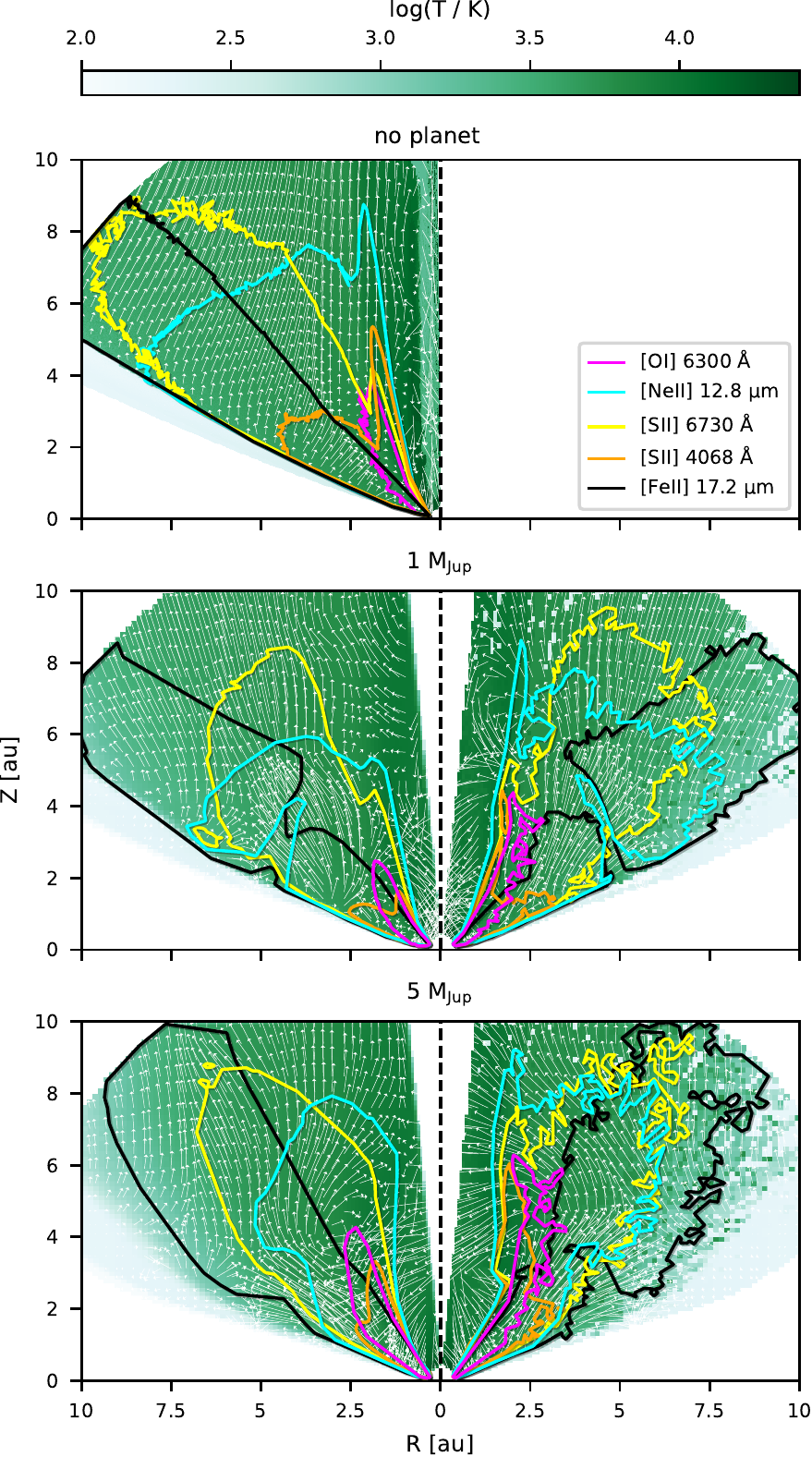}
    \caption{
        85\% emission regions of our chosen diagnostic lines. The left halves show the azimuthal average, the right halves show a cut through the y=0 plane, where the planet is located. The background colour indicates the temperature that was calculated with MOCASSIN. The white arrows indicate the velocity field of the wind.
    }
    \label{fig:emission-maps}
\end{figure}

One of the most important diagnostic tools that exist for analysing protoplanetary disc winds are optical forbidden emission lines. These collisionally excited lines are particularly well suited to study disc winds, because they depend strongly on the temperature through the Boltzmann term exp(-$\Delta$k$_\mathrm{B}$T), but also on the density. In the bound protoplanetary disc, the temperature is usually low and the density high enough for collisional de-excitation to dominate, which means that the forbidden line emission is suppressed in the disc itself. However, disc-winds can have the right conditions to allow for significant emission. The ideal conditions vary from line to line, opening up the possibility to study different regions of a wind by analysing different emission lines.
Figure \ref{fig:emission-maps} shows the temperature that is resulting from our photo-ionisation calculations of the wind models as well as the 85\% emission regions of the lines in our sample, namely the optical [OI]~6300~\angstrom, [SII]~4068\angstrom{} and [SII]~6730~\angstrom lines and in the mid-infrared the [NeII]~12.8~\umt fine structure emission line and the [FeII]~17.2~\umt line. When the model includes a planet the left panel shows the azimuthal average and the right panel the emission region in the y=0 plane, where the planet is situated. We include this comparison in order to illustrate the typical magnitude of the deviation from the mean, which we expect to be strongest at the azimuthal position of the planet. Generally the averaged emission regions are in good agreement with the emission in the y=0 plane, although in the latter all lines show a higher vertical extent along the inner edge and both show reduced emission inside the under-density above the planetary gap, but the void that this creates in the emission region is larger in the y=0 plane, which is expected, as asymmetries within the disc would wash out the feature in the average. The low emission inside this region can be interpreted as a consequence of less efficient collisional excitation at lower densities. It is possible that this could to a degree be compensated when material accretes onto the planet and forms strong shocks, emitting additional EUV-radiation that could reach and heat the wind. In that case the forbidden lines would trace the under-density better and the effect on the line profile would be even stronger. We note that the emission regions show the contours within which 85\% of the emission is originating and do not indicate the absolute luminosity. This means that the two phenomena, the reduced emission inside the under-density and the increased vertical extent, are likely to be linked, as taking away emission from one part of the wind will result in the 85\% contours being shifted towards another part. 

Of the lines in our sample only the [SII]~6730~\angstrom{}, the [NeII]~12.8~\umt and the [FeII]~17.2~\umt lines reach far enough into the wind to cover the region above the gas cavity, however, as the emission is strongly reduced inside the under-density, these lines do not trace that region very well, either. Nevertheless, as the velocity structure of the wind is modified far beyond the under-density, even reaching to the innermost parts of the wind, all lines do trace parts of the wind that are in some way affected by the planet. At larger radii, the models that include a planet experience a smoother transition between low and high temperatures at the disc-wind interface outside the cavity. As a result, the line emission is suppressed close to the disc surface, and the emission region pushed to higher heights above the disc surface. The [FeII]~17.2~\umt line is the least affected by this, such that in model MJ1 its 85\% emission region reaches down to the disc surface. It is also the line that with least emission along the inner edge of the wind, making it a promising diagnostic line for tracing the actual launching region of extended disc winds.

\subsubsection{Line profiles} \label{sec:res:profiles}
\begin{figure*}
    \centering
    \includegraphics[width=.8\textwidth]{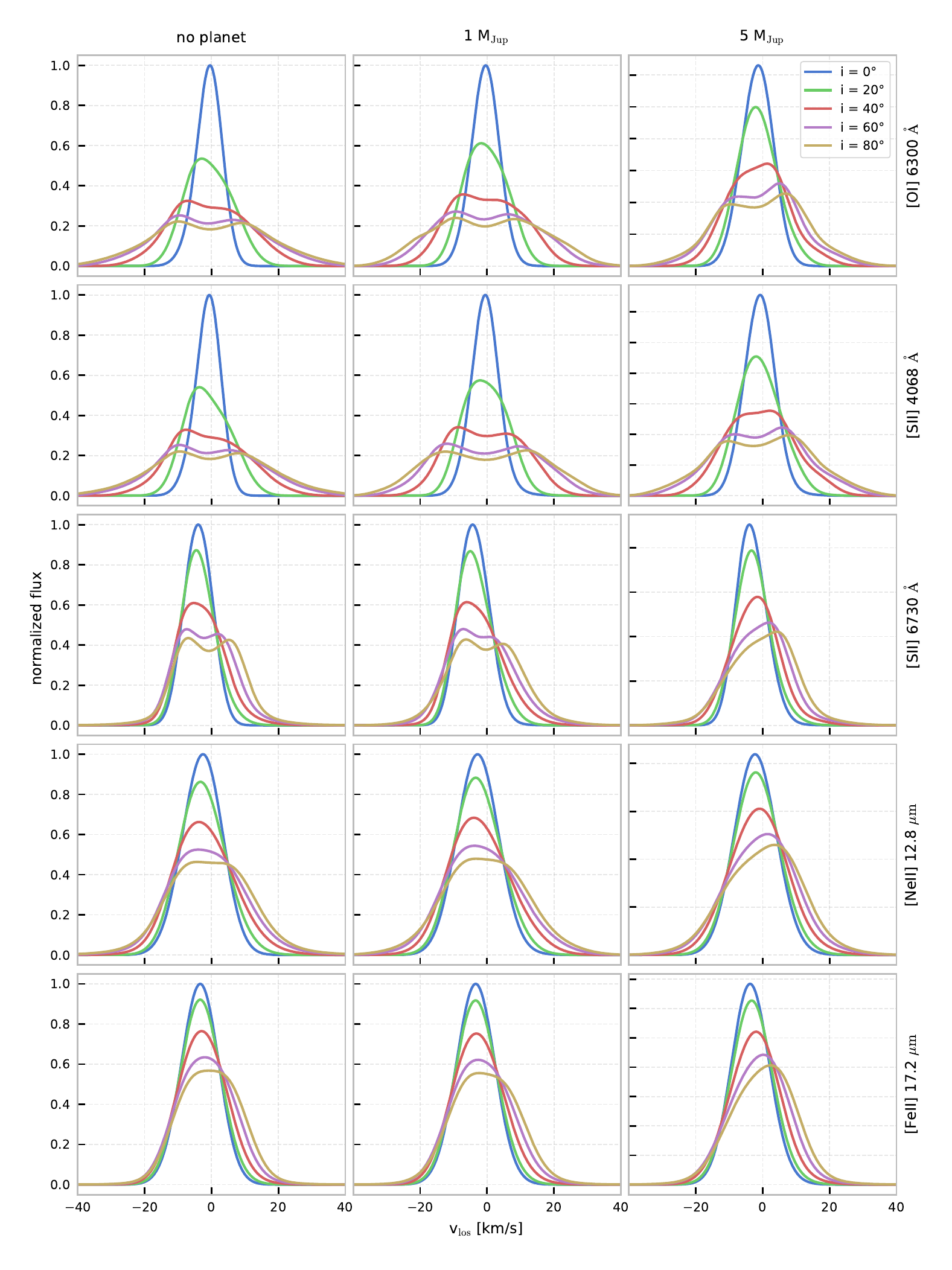}
    \caption{
        Synthetic spectral profiles of the optical lines viewed at inclination i and azimuth angle $\phi = 0$. The profiles were artificially degraded to a spectral resolution corresponding to R = 100000 (optical) and R = 60000 (mid-infrared) and normalised to the peak of the face-on profile with i = 0\degree. 
    }
    \label{fig:profiles-all-future-res}
\end{figure*}

To study the structure and kinematics of protoplanetary disc winds it is useful to analyse the spectral line profiles of the emission lines. We have calculated artificial line profiles from our models as described in section \ref{sec:met:profiles} and convolved them with a Gaussian function, in order to simulate a finite spectral resolution that would be the limiting factor in real observations. Figure \ref{fig:profiles-all-future-res} shows the resulting profiles. The line profiles are artificially degraded to a resolving power of R = 100000 and R = 60000 for the optical lines and mid-infrared lines, respectively.
We show profiles with half the resolution in appendix \ref{sec:appendix:profiles}.
Since the line profiles always trace an extended region of the disc wind that covers a range of wind velocities, changes in the flow structure in one region of the wind are not easily distinguishable from a change of the luminosity in a different region of the wind. In order to be able to observe a difference between models, the affected region must either be very large, or very luminous with a drastically different flow structure. 

In our model MJ1 the wind is affected by the gas-gap only to a limited extent. Although the emission regions extend over wind regions where the outflow is affected, no significant difference in the line profiles can be seen, when comparing them to the reference model without a planet. When not observed face-on, the [OI]~6300~\angstrom line and the very similar [SII]~4068~\angstrom line are slightly less blueshifted, but it would be difficult to draw any conclusions from such a small effect. As an example, the Gaussian fit for the [OI]~6300~\angstrom profile observed at i=40\degree and $\Phi$=0\degree yields a centroid velocity of -0.66~km/s compared to -1.16~km/s in the reference model (see table $\ref{tab:fits_oi-6300}$ in the appendix). Since the majority of the line emission originates from close-in, the contribution to the line profile that traces parts of the wind that are the most affected by the presence of a planet is small. 

In the MJ5 model, however, the perturbations of the flow structure are much stronger, as already seen in Figure \ref{fig:wind-structure}. Not only is the gap and with it the affected volume of the wind much larger, but also the wind-launching region in the inner disc is affected much more. The effects are sufficiently strong to make an impact on the line profiles. The profiles show less blueshift when observed at small inclinations while at higher inclinations the peak is shifted into the redshifted part of the spectrum. All of the modelled optical lines have their peak at velocities > 0 when observed at inclinations of 60\degree or more.
Although the redshift is small (only few km/s), it is a notable observation, as it requires a significant portion of the wind to move away from the observer. Such a scenario cannot easily be explained by a simple wind model alone. Even in transition disc models, where a portion of the receding wind on the other half of the disc can be observed through the large inner gas cavity, the profiles only show enhanced symmetry and less blueshift, but cannot exhibit a redshifted peak \citep{Picogna2019}. 

The [SII]~6730~\angstrom line is the only optical line in our model that features a single peak even at high inclinations. With its large emission region, it's profile is not dominated by the regions close to the star that are subject to strong Keplerian broadening. Instead, it traces much of the wind-region that is most affected by the planetary gap. As a result the line is strongly skewed towards the red with a peak in the red part of the spectrum at i$\geq$60\degree. The [OI]~6300~\angstrom and [SII]~4068~\angstrom lines both start to exhibit a redshifted peak already at i=40°. At higher inclinations, their profiles show clear Keplerian double peaks with the dominant peak in the red. This indicates that the red part of the profiles not only traces the region above the gap, but the presence of a massive planet farther out also affects the inner parts of the wind, as was already discussed in section \ref{sec:res:structure}.
The lines in the mid-infrared trace a similarly extended region as [SII]~6730~\angstrom, but are degraded to a lower spectral resolution. Nonetheless, they behave very similarly with redshifted peaks at i$\geq$60\degree ([NeII]~12.8~\umt) and i$\geq$80\degree ([FeII]~17-2~\umt).

In a typical analysis, where the profiles are decomposed into multiple Gaussian components, the features described above would easily be missed. Because the profiles are skewed or exhibit a Keplerian double-peak feature, it is clear that with increasing spectral resolution, most are not well represented by Gaussian components, especially when observed at intermediate to high inclinations. With upcoming instruments that will allow for observations with higher resolution, it becomes worthwhile to study the detailed shape of the profiles. To show this, we list the results of the Gaussian decomposition of the [OI]~6300~\angstrom line profiles as well as the velocity of the actual peak in table \ref{tab:fits_oi-6300}. The results for the other lines are available as online supplementary material. When the decomposition yields only a single component, the centroid velocity is almost always in the blueshifted part of the spectrum, even if the peak is clearly redshifted. At higher inclinations, starting at $\approx40\degree$, the decomposition often yields two narrow components that fit to the Keplerian double peaks. Such a decomposition does not allow for a clear identification of a redshifted peak either, as it often fails to produce a fit that accurately matches the peaks of the high-resolution profiles.

For reference, we also show profiles calculated with half the spectral resolution, comparable to currently available observations, in appendix \ref{sec:appendix:profiles}. It is clear that the effects of the planet on the wind cannot be unambiguously observed in currently available spectra. Forbidden line diagnostics of disc winds will greatly benefit from the anticipated high resolution spectra.

\subsubsection{Profile variability} \label{res:profile-variability}
\begin{figure}
    \centering
    \includegraphics[width=.48\textwidth]{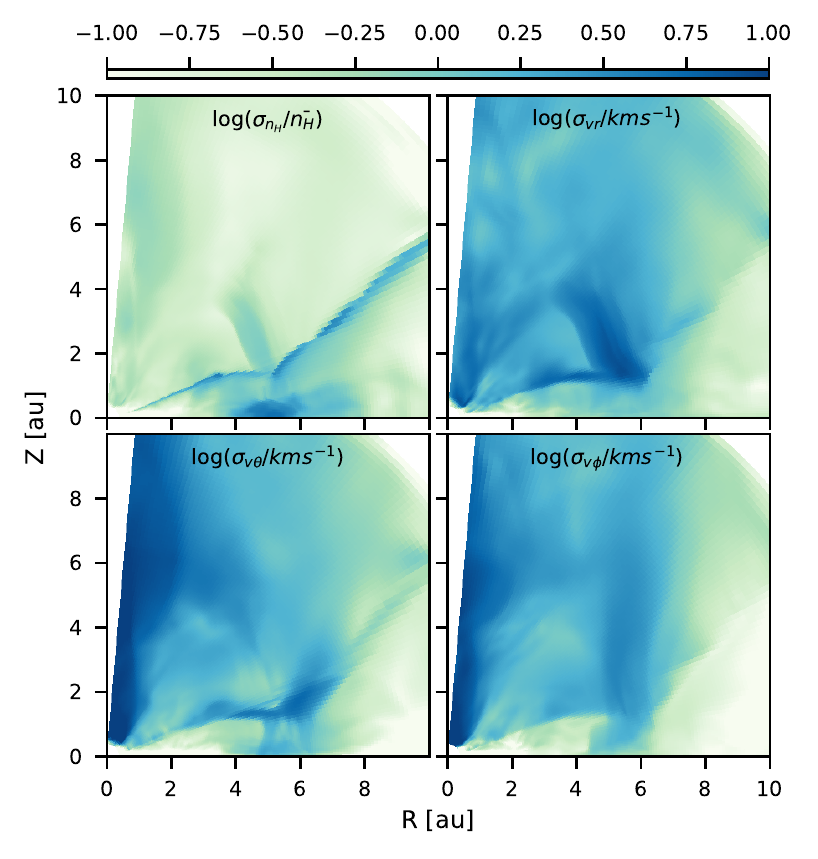}
    \caption{
        Maps of the standard deviation $\sigma$ along the azimuthal direction for the hydrogen number density and the three velocity components in spherical coordinates. For the number density, the map is weighted by the local azimuthal mean $\bar{n_H}$.
    }
    \label{fig:asymmetry}
\end{figure}

\citet{Simon2016a} compared their high-resolution ($\sim$6~km/s) [OI]~6300~\angstrom profiles to the slightly lower-resolution ($\sim$12~km/s) profiles that were observed earlier by \citet{Hartigan1995}. While they found that the general structure of the LVC remains mostly stable over the timescale of one or two decades, they did find objects that had variations between the different epochs. Two of the objects in their high-resolution sample, UX Tau A and IP Tau, were observed twice, $\sim$6~years apart. The profile of IP Tau evolved so strongly that the LVC vanished from the decomposition of the later profile. UX Tau A experienced significant variations, too.

\begin{figure*}
    \centering
    \includegraphics[width=.65\textwidth]{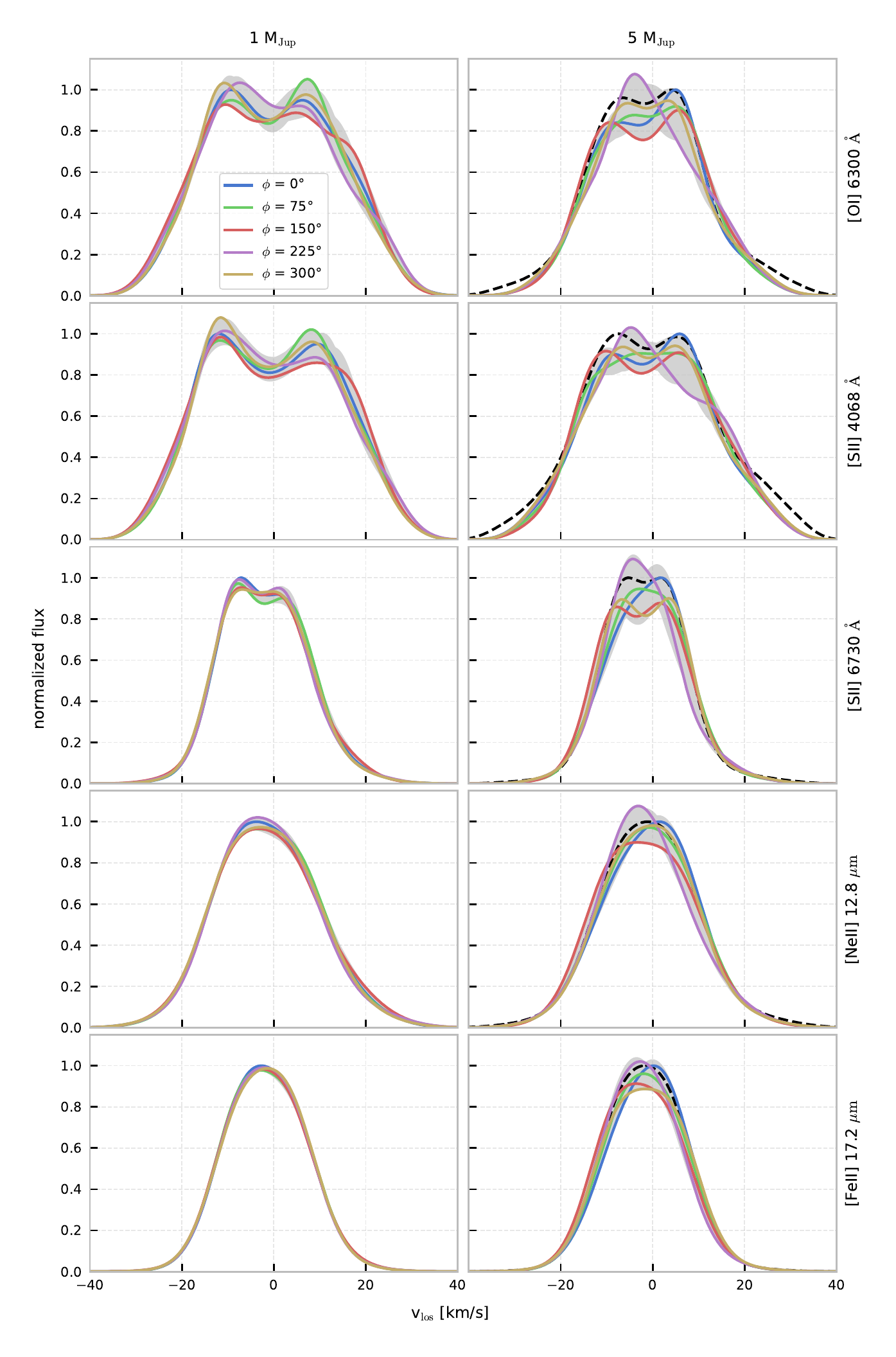}
    \caption{
        Synthetic spectral profiles viewed at 60\degree inclination but at different azimuth angles $\phi$. The grey shaded area encloses the minimum and maximum flux at a certain velocity that can be observed during one orbit of the planet. The profiles were degraded to a resolving power of R = 100000 and normalised to the peak flux of the $\phi = 0$ profile.
    }
    \label{fig:profile-variability}
\end{figure*}

In our three-dimensional models, the substructures such as the spiral waves and instabilities that are induced by the planet in the disc do have an impact on the azimuthal symmetry of the wind. We show the extent of the asymmetries in Figure \ref{fig:asymmetry}, were we plot a map of the standard deviation along the azimuth. Since the number density spans a large range of values of several orders of magnitude, we weighted its standard deviation with the local azimuthal mean value of the number density. The deviation inside the wind reaches up to $\sim$30\% of the mean. The velocity components show a deviation of several km/s. This is sufficient to lead to variations in our synthetic line profiles on timescales shorter than the planet's orbit. Figure \ref{fig:profile-variability} shows the variations that would be observed with a high spectral resolution (R = 100000) when the disc is observed at an inclination of 60\degree and from different azimuth angles. 
This crude approximation of the variations that would be observed during an orbit of the planet neglects the wind dynamics. The resulting variations should therefore be considered a lower limit. However, we expect the influence of the wind dynamics to be small, because the hydrodynamical timescales over which variations in the wind structure propagate far enough to cover a significant portion of the line emission region are of the order of years (at a typical speed of 5~km/s high up in the wind, it takes $\approx1$ year to travel 1~au).

Variability is observable in the [OI]~6300~\angstrom and [SII]~4068~\angstrom profiles of the MJ1 model and it is even more evident in the MJ5 model, which shows significant variations in all other lines, too. The profiles experience observable shifts of the peak velocity on a timescale of less than a quarter of the planet's orbital period, which corresponds to $\approx$3.5~years in our models. Around the location of the peak the flux varies by 10 -- 20 \%. A signal to noise ratio between 25 and 50 should therefore be sufficient to detect the variability at the simulated spectral resolution.
For model MJ5 we also show synthetic profiles that were calculated from a time-average of the hydrodynamical grid taken from 850 snapshots over $\approx$5~orbits of the planet. This is to show that the presented profiles, which are individual snapshots, are reasonably scattered around the average and are not an edge-case. 

In the MJ5 model the average shows that the redshifted peak is dominating most of the time in the [OI]~6300~\angstrom line profile, but not in the profiles of the other lines. Given that the overall wind structure is in an approximately steady state with only minor dynamical fluctuations over multiple orbits, the variations shown here should be periodic in nature with a period that increases with the planet's separation. Due to the high computational cost of the simulations, however, we did not explore models with different separations.

One caveat may be that the planet-induced variability could be overshadowed by variability of the stellar accretion luminosity that is mainly responsible for the heating of the wind and drives the emission of our modelled lines. As was shown e.g. by \citet{Weber2020}, the line profile properties correlate with the accretion luminosity. Although they find that this correlation is strongest for the line luminosity, they do find that it correlates with the width and centroid-velocity of the fitted Gaussian components and can slightly affect the line profile shape. Even though the effect on the shape of the line profile appears small, it could complicate the interpretation of profile variations, especially since variability of the accretion luminosity has been observed over much shorter timescales on the order of days \citep{Manara2021}.

\subsubsection{Comparison with observations} \label{res:comparison}
\begin{figure*}
    \centering
    \includegraphics[width=\textwidth]{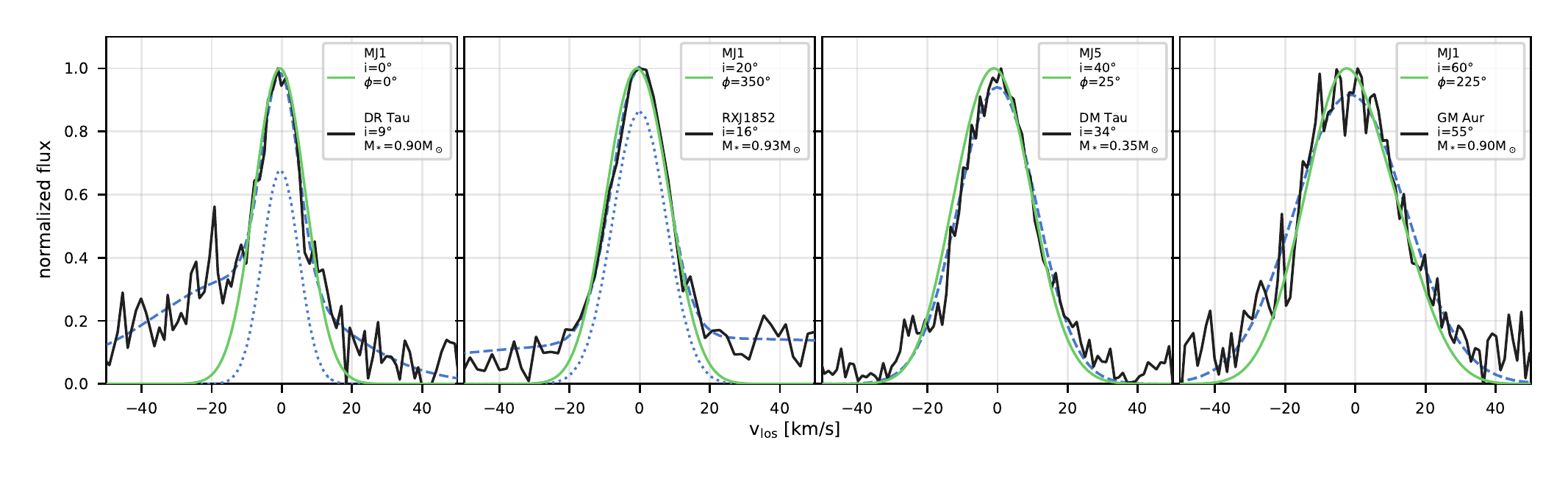}
    \caption{
        Comparison of our synthetic [OI]~6300~\angstrom profiles with selected observations. The black solid line shows the observed line profiles taken from \citet{Banzatti2018}. The inclinations and stellar masses are taken from \citet[][table 1]{Banzatti2018}. The blue dashed line shows the fit to the observed profile, the blue dotted line shows the NC of the fit. The green solid line shows the modelled profile that matches best between all profiles from all three models.
    }
    \label{fig:observations}
\end{figure*}

In Figure \ref{fig:observations} we show for inclinations of 0\degree, 20\degree, 40\degree and 60\degree a well-matching pair of a modelled and an observed [OI]~6300~\angstrom profile. The observations were selected from the combined observational sample of \citet{Fang2018} and \citet{Banzatti2018}. In these works, the authors also list references for the inclinations of most objects, which we used to limit the comparisons to objects with an inclination that deviates not more than 10\degree from that of the model. Due to the lack of near edge-on observations, we do not include a comparison at 80\degree. We use the [OI]~6300~\angstrom line, because it is the line with the most available observations. To find a good match, we degraded the synthetic profiles to a resolving power R = 45000, comparable to that of the observations and then chose the model-observation pair that minimises $\chi^2$ in the intervals [-15, 15], [-19, 19], [-23, 23] and [-26, 26] km/s for i=0\degree, 20\degree 40\degree and 60\degree, respectively. We extend the intervals with increasing inclination in order to account for the broadening that is expected due to Keplerian rotation and we ignore everything outside the intervals, because our models can reproduce only narrow low-velocity components and not the broad or high-velocity components that are often additionally observed and usually attributed to magnetic winds. We note that this method will favour observations that have a distinct, isolated narrow component. However, it is not possible to unambiguously extract the narrow component from more complex line profiles that are likely to contain additional emission that traces a magnetic wind. Moreover, the noise of the observed profiles is often larger than the difference between our synthetic profiles at the available resolution. With the currently available data our approach is therefore not suitable for a quantitative comparison between different models, or to distinguish between models with and without planets, as for example also a model of a photo-evaporating transition disc without a planet but with a completely depleted inner gas cavity typically leads to profiles with lower blueshifts. Nevertheless, we consider this comparison to be a good demonstration that our models can reproduce the often observed narrow low-velocity components very well.

The resulting matches are all between one of the models that contains a planet and an object with a clear narrow component in it's line profile that is only slightly blueshifted ($v_c$ = -0.16, -0.21, -0.16 and -1.77 for DR~Tau, RX~J1852, DM~Tau and GM~Aur, respectively). Due to the limited resolution, neither the models nor the observations exhibit any Keplerian double peaks or obviously skewed profiles where a single redshifted peak could be identified. At that resolution, profiles with similarly low blueshifts could possibly be also reproduced with a model of a photo-evaporating transition disc without a planet, as a completely depleted inner gas cavity typically leads to profiles with lower blueshifts \citep{Picogna2019}. However, it is remarkable that although none of our models have been tuned to reproduce any of the observations, they match the observations very well across all available inclinations and across different types of discs. DR Tau is considered a very active classical T~Tauri star with a complex line profile that can be decomposed into high-velocity and broad and narrow low-velocity components. \citet{Mesa2022} detected a compact and elongated structure in its disc at a separation of 303$\pm$10~mas, corresponding to $\approx59$~au, which they suggest could hint at the presence of an embedded forming planet with a few M$_\mathrm{J}$ in mass. In contrast, RX~J1852, DM~Tau and GM~Aur, are all known accreting transition discs with simpler line profiles and with observed gaps that could be caused by planets. DM~Tau and GM~Aur have been particularly well studied. \citet{Bosman2021} found evidence for a strongly depleted gas cavity inside R~$\approx 15$~au in the disc around GM~Aur. \citet{Kudo2018} detected a dust ring at R~$\approx 4$~au in the 1.3mm dust continuum emission and $^{12}$CO~J~=~2~$\xrightarrow{}$~1 emission of DM~Tau that could potentially indicate the presence of a planetary gap at a location that is similar to that in our models. But even then a direct comparison to our model would not be possible, because DM~Tau has only half the stellar mass of the star in our models. We will discuss in the following section how our results may be affected by different model parameters such as the location of the planetary gap and the stellar mass.

\subsection{Dependence on model parameters} \label{res:parameter-dependence}

Although the high computational cost of the simulations prevent us from doing an extensive parameter study, we can nevertheless apply theoretical considerations to get insight on the influence of the key parameters in our models. \citet{Kanagawa2016} found a relation for the gap width $\Delta_{gap}$:
\begin{equation} \label{eq:gap-width}
    \frac{\Delta_{gap}}{a_p} = 0.41 \bigg(\frac{M_p}{M_*}\bigg)^{\frac{1}{2}} \bigg(\frac{h_p}{a_p}\bigg)^{-\frac{3}{2}} \alpha^{-\frac{1}{4}},
\end{equation}
where $a_p$ is the orbital radius of the planet, $h_p$ the scale height of the disc at the location of the planet and $\alpha$ is the viscosity parameter. Since the width of the gap determines the size of the affected wind-region, we can use this to investigate how our results depend on an individual parameter, while keeping the others constant. 

\subsubsection{Dependence on the planetary mass} \label{sec:dis:params:stellar-mass}
We have already seen in section \ref{sec:res:structure} that an increase in planet mass leads to a wider gap, which leads to stronger effects on the wind structure and on the line profiles. We have also shown that the gravitational potential of the planet is negligible in the wind regions. From relation \eqref{eq:gap-width} we would therefore expect the gap width to increase by a factor $\sqrt{5} \approx 2.24$ between the models MJ5 and MJ1. When measured between the points at the gap edges, where the azimuthally averaged 1D surface density profile reaches half of the unperturbed value at the location of the planet, we find a factor 2.69 increase in our models, which is in reasonably good agreement.

\subsubsection{Dependence on the stellar mass}
According to equation $\eqref{eq:gap-width}$, the dependence of the gap-width on the stellar mass is inverse to that of the planet. The combination of a 0.14~M$_\odot$ star with a Jupiter-like planet should give a similar gap width as we find in model MJ5. However the stellar mass has a strong impact on the structure of the photo-evaporative wind, because it not only defines the gravitational potential that the wind has to overcome, but it also affects the irradiating spectrum and luminosity that is driving the wind \citep{Ercolano2021, Picogna2021}. A lower stellar mass allows the wind to be launched at radii closer to the star, but it is also less extended. In such a scenario it is possible that the EUV radiation that heats the wind and drives the forbidden line emission will to some degree be screened by the inner wind and not reach the region that is the most affected by the presence of the planet. In that case the distinctive features in the line profiles would be weaker, but detailed models are required in order to determine how the lower wind-launching radius affects the wind dynamics and the forbidden emission lines.

\subsubsection{Dependence on the viscosity}
We do not expect the viscosity to have a significant effect on the dynamics of the wind, however since the gap width scales with $\alpha^{-\frac{1}{4}}$, we should be able to reproduce effects similar to that of increasing the planet mass, when we decrease the viscosity, instead. As an example, a model with $\alpha = 4\cdot10^{-5}$ and M$_p$ = M$_J$ should be comparable to our model MJ5. 

\subsubsection{Dependence on the orbital radius}
Assuming a constant aspect ratio of the disc, if the orbital radius of the planet is larger, the gap will be proportionally wider (see eq. \eqref{eq:gap-width}). However since the forbidden line emission is generally dominated in the wind-regions close to the star, most lines will not be able to trace the regions above the gap if the planet is outside $\approx10$~au. Although our models show that the presence of the gap modifies the wind dynamics even in the inner parts of the disc, it is unclear whether this would happen with a gap at a larger orbital radius. Because the photo-evaporative wind is weaker at larger radii, the pressure difference between the under-density and the wind at the inner gap-edge will be smaller, too, which will lead to smaller horizontal components in the pressure gradient.
A lower orbital radius would shift the gap closer to the star and closer to the wind-launching region that dominates the emission. In that case, we expect the effect on the line-profiles to become stronger and the timescales of the profile-variability to become shorter, as long as the gap does not reach the critical launching radius of the wind ($\approx 2$~au, see Figure \ref{fig:wind-structure}). When measuring the width of the gap as described in section \ref{sec:dis:params:stellar-mass}, we can get the estimation from Figure \ref{fig:wind-structure} that no wind is launched inside the inner $\approx$75~\% of the gap. Assuming that this ratio is constant, we find with eq. \eqref{eq:gap-width} an estimated threshold orbital radius of $\approx2.5$~au and $\approx3.3$~au for M$_p$ = M$_J$ and M$_p$ = 5~M$_J$, respectively. If the orbital radius was lower than the threshold, a photo-evaporative wind could only be launched outside the gap. The emission would then likely shift outwards, too, which would lead to narrower lines and less variability, due to the slower Keplerian rotation at higher radii.

\subsubsection{Dependence on the X-ray luminosity}
The X-ray luminosity is the main driver of the photo-evaporative wind in our model. A higher X-ray luminosity leads to a more vigorous wind with a higher total mass loss. With a denser wind we expect the contrast between the under-density and the wind to be stronger, leading to stronger pressure gradients, which could amplify the features that we observed in the line profiles of the MJ5 model. In contrast to that, while \citet{Ercolano2016} have shown that there is no significant correlation between the luminosity of forbidden lines and the X-ray luminosity, we do expect the emission regions to be less extended when the wind is denser, because the EUV-radiation is more efficiently absorbed. Again, detailed models would be required to quantify these competing effects.

\subsubsection{Dependence on the accretion luminosity}
Although the accretion luminosity has no significant impact on the disc or wind structure, its EUV component is the main driver for the line emission \citep{Ercolano2016}. \citet{Weber2020} have shown that when the accretion luminosity is higher, the size of the emission regions increases. That way, the forbidden lines could potentially trace more of the region that is affected by the presence of the planet, leading to stronger features in the line profiles with higher accretion luminosity.


\section{Conclusions}\label{sec:conclusions}
We have created three-dimensional numerical models of photo-evaporating, viscous ($\alpha = 10^{-3}$) protoplanetary discs hosting a giant planet with the goal of studying the interplay between planet-disc interactions and photo-evaporation. In order to predict the imprint on forbidden emission line diagnostics, we have performed detailed photo-ionisation and radiative transfer simulations and created synthetic spectral line profiles from the results. The models show that
\begin{itemize}
    \item A massive, gap-opening planet can significantly alter the photo-evaporative wind structure at and inside the radial location of the gap. Above the gap, the pressure gradient that drives the wind is inversed, pointing downwards. As a result, a region of reduced density develops above the gap and wind-material that enters this region from the inner disc wind or from above the outer gap-edge can fall back down and into the gap. It is unclear whether winds that are not driven by a pressure gradient (e.g. magnetic winds) would experience similar effects.
    
    \item A gap carved by a 5~M$_J$ planet can show an imprint on the spectral line profiles of forbidden line emission in high resolution (R~$ \approx100000$) spectra. In all of our synthetic emission lines the peak of the profile could be shifted to the red part of the spectrum by several km/s, when the disc is observed at an inclination of 40\degree -- 60\degree or more. These details are lost in a multi-Gaussian decomposition, demonstrating the need for more detailed analyses of line profiles with increasing spectral resolution. A much less massive planet (1 M$_\mathrm{J}$) cannot affect the wind-structure enough to have an observable effect on the line profiles, unless the viscosity in the disc is much lower than in our models. 
    
    \item Asymmetric substructures that are generated by the planet within the disc leave a signature in the photo-evaporative wind. This gives rise to temporal variations of the line profile shape that can be strong enough to be observable on timescales of less than a quarter of the planet's orbital period.
    
    \item Our synthetic line profiles compare well to a subset of observations of objects with different inclinations and stellar masses. However, the resolution of currently available observations is not sufficient to allow for an identification of the distinctive features that our models would predict in the presence of a planetary gap. Future observations with about twice the currently available spectral resolution could provide enough detail for this to become possible.
\end{itemize}

\section*{Acknowledgements}\label{sec:acknowledgements}
We are grateful to the anonymous referee, whose comments helped improve this work. This research was supported by the Excellence Cluster ORIGINS which is funded by the Deutsche Forschungsgemeinschaft (DFG, German Research Foundation) under Germany´s Excellence Strategy – EXC-2094-390783311. BE, GP and CHR acknowledge the support of the Deutsche Forschungsgemeinschaft (DFG, German Research Foundation) - 325594231. CHR is grateful for support from the Max Planck Society. The simulations have been partly carried out on the computing facilities of the Computational Center for Particle and Astrophysics (C2PAP).

\section*{Data Availability}
The data underlying this article are available from the authors upon request.


\bibliographystyle{mnras}
\bibliography{references}

\begin{thebibliography}{}
\makeatletter
\relax
\def\mn@urlcharsother{\let\do\@makeother \do\$\do\&\do\#\do\^\do\_\do\%\do\~}
\def\mn@doi{\begingroup\mn@urlcharsother \@ifnextchar [ {\mn@doi@}
  {\mn@doi@[]}}
\def\mn@doi@[#1]#2{\def\@tempa{#1}\ifx\@tempa\@empty \href
  {http://dx.doi.org/#2} {doi:#2}\else \href {http://dx.doi.org/#2} {#1}\fi
  \endgroup}
\def\mn@eprint#1#2{\mn@eprint@#1:#2::\@nil}
\def\mn@eprint@arXiv#1{\href {http://arxiv.org/abs/#1} {{\tt arXiv:#1}}}
\def\mn@eprint@dblp#1{\href {http://dblp.uni-trier.de/rec/bibtex/#1.xml}
  {dblp:#1}}
\def\mn@eprint@#1:#2:#3:#4\@nil{\def\@tempa {#1}\def\@tempb {#2}\def\@tempc
  {#3}\ifx \@tempc \@empty \let \@tempc \@tempb \let \@tempb \@tempa \fi \ifx
  \@tempb \@empty \def\@tempb {arXiv}\fi \@ifundefined
  {mn@eprint@\@tempb}{\@tempb:\@tempc}{\expandafter \expandafter \csname
  mn@eprint@\@tempb\endcsname \expandafter{\@tempc}}}

\bibitem[\protect\citeauthoryear{{ALMA Partnership} et~al.,}{{ALMA Partnership}
  et~al.}{2015}]{ALMAPartnership2015}
{ALMA Partnership} et~al., 2015, \mn@doi [Astrophysical Journal Letters]
  {10.1088/2041-8205/808/1/L3}, 808, L3

\bibitem[\protect\citeauthoryear{Alexander}{Alexander}{2008}]{Alexander2008}
Alexander R.~D.,  2008, \mn@doi [Monthly Notices of the Royal Astronomical
  Society: Letters] {10.1111/j.1745-3933.2008.00556.x}, 391, L64

\bibitem[\protect\citeauthoryear{Alexander \& Armitage}{Alexander \&
  Armitage}{2009}]{Alexander2009}
Alexander R.~D.,  Armitage P.~J.,  2009, \mn@doi [Astrophysical Journal]
  {10.1088/0004-637X/704/2/989}, 704, 989

\bibitem[\protect\citeauthoryear{Andrews et~al.,}{Andrews
  et~al.}{2018}]{Andrews2018}
Andrews S.~M.,  et~al., 2018, \mn@doi [The Astrophysical Journal]
  {10.3847/2041-8213/aaf741}, 869, L41

\bibitem[\protect\citeauthoryear{Bai}{Bai}{2015}]{Bai2015}
Bai X.~N.,  2015, \mn@doi [Astrophysical Journal] {10.1088/0004-637X/798/2/84},
  798, 84

\bibitem[\protect\citeauthoryear{Bai}{Bai}{2017}]{Bai2017}
Bai X.-N.,  2017, \mn@doi [The Astrophysical Journal]
  {10.3847/1538-4357/aa7dda}, 845, 75

\bibitem[\protect\citeauthoryear{Balbus \& Hawley}{Balbus \&
  Hawley}{1991}]{Balbus1991}
Balbus S.~A.,  Hawley J.~F.,  1991, \mn@doi [The Astrophysical Journal]
  {10.1086/170270}, 376, 214

\bibitem[\protect\citeauthoryear{Ballabio, Alexander  \& Clarke}{Ballabio
  et~al.}{2020}]{Ballabio2020}
Ballabio G.,  Alexander R.~D.,   Clarke C.~J.,  2020, \mn@doi [Monthly Notices
  of the Royal Astronomical Society] {10.1093/MNRAS/STAA1767}, 496, 2932

\bibitem[\protect\citeauthoryear{Banzatti, Pascucci, Edwards, Fang, Gorti  \&
  Flock}{Banzatti et~al.}{2019}]{Banzatti2018}
Banzatti A.,  Pascucci I.,  Edwards S.,  Fang M.,  Gorti U.,   Flock M.,  2019,
  \mn@doi [The Astrophysical Journal] {10.3847/1538-4357/aaf1aa}, 870, 76

\bibitem[\protect\citeauthoryear{Bast, Brown, Herczeg, van Dishoeck  \&
  Pontoppidan}{Bast et~al.}{2011}]{Bast2011}
Bast J.~E.,  Brown J.~M.,  Herczeg G.~J.,  van Dishoeck E.~F.,   Pontoppidan
  K.~M.,  2011, \mn@doi [Astronomy {\&}amp; Astrophysics, Volume 527, id.A119,
  <NUMPAGES>18</NUMPAGES> pp.] {10.1051/0004-6361/201015225}, 527, A119

\bibitem[\protect\citeauthoryear{B{\'{e}}thune, Lesur  \&
  Ferreira}{B{\'{e}}thune et~al.}{2017}]{Bethune2017}
B{\'{e}}thune W.,  Lesur G.,   Ferreira J.,  2017, \mn@doi [Astronomy and
  Astrophysics] {10.1051/0004-6361/201630056}, 600, A75

\bibitem[\protect\citeauthoryear{Bosman et~al.,}{Bosman
  et~al.}{2021}]{Bosman2021}
Bosman A.~D.,  et~al., 2021, \mn@doi [The Astrophysical Journal Supplement
  Series] {10.3847/1538-4365/ac1433}, 257, 15

\bibitem[\protect\citeauthoryear{De~Val-Borro et~al.,}{De~Val-Borro
  et~al.}{2006}]{DeVal-Borro2006}
De~Val-Borro M.,  et~al., 2006, \mn@doi [Monthly Notices of the Royal
  Astronomical Society] {10.1111/j.1365-2966.2006.10488.x}, 370, 529

\bibitem[\protect\citeauthoryear{Dere, Landi, Mason, Monsignori~Fossi  \&
  Young}{Dere et~al.}{1997}]{Dere1997}
Dere K.~P.,  Landi E.,  Mason H.~E.,  Monsignori~Fossi B.~C.,   Young P.~R.,
  1997, \mn@doi [Astronomy and Astrophysics Supplement Series]
  {10.1051/aas:1997368}, 125, 149

\bibitem[\protect\citeauthoryear{Dere, Zanna, Young, Landi  \& Sutherland}{Dere
  et~al.}{2019}]{Dere2019}
Dere K.~P.,  Zanna G.~D.,  Young P.~R.,  Landi E.,   Sutherland R.~S.,  2019,
  \mn@doi [The Astrophysical Journal Supplement Series]
  {10.3847/1538-4365/ab05cf}, 241, 22

\bibitem[\protect\citeauthoryear{Dullemond, Hollenbach, Kamp  \&
  D'Alessio}{Dullemond et~al.}{2006}]{Dullemond2007}
Dullemond C.~P.,  Hollenbach D.,  Kamp I.,   D'Alessio P.,  2006, in Reipurth
  B.,  Jewitt D.,   Keil K.,  eds, Protostars and Planets V. p.~555, \url
  {http://arxiv.org/abs/astro-ph/0602619}

\bibitem[\protect\citeauthoryear{D’Alessio, Calvet  \& Hartmann}{D’Alessio
  et~al.}{2001}]{DAlessio2001}
D’Alessio P.,  Calvet N.,   Hartmann L.,  2001, \mn@doi [The Astrophysical
  Journal] {10.1086/320655}, 553, 321

\bibitem[\protect\citeauthoryear{Engvold}{Engvold}{1977}]{Asplund2005}
Engvold O.,  1977, \mn@doi [Physica Scripta] {10.1088/0031-8949/16/1-2/007},
  16, 48

\bibitem[\protect\citeauthoryear{Ercolano \& Owen}{Ercolano \&
  Owen}{2010}]{Ercolano2010}
Ercolano B.,  Owen J.~E.,  2010, \mn@doi [Monthly Notices of the Royal
  Astronomical Society] {10.1111/j.1365-2966.2010.16798.x}, 406, 1553

\bibitem[\protect\citeauthoryear{Ercolano \& Owen}{Ercolano \&
  Owen}{2016}]{Ercolano2016}
Ercolano B.,  Owen J.~E.,  2016, \mn@doi [Monthly Notices of the Royal
  Astronomical Society] {10.1093/mnras/stw1179}, 460, 3472

\bibitem[\protect\citeauthoryear{Ercolano \& Pascucci}{Ercolano \&
  Pascucci}{2017}]{Ercolano2017a}
Ercolano B.,  Pascucci I.,  2017, \mn@doi [Royal Society Open Science]
  {10.1098/rsos.170114}, 4

\bibitem[\protect\citeauthoryear{Ercolano \& Rosotti}{Ercolano \&
  Rosotti}{2015}]{Ercolano2015a}
Ercolano B.,  Rosotti G.,  2015, \mn@doi [Monthly Notices of the Royal
  Astronomical Society] {10.1093/mnras/stv833}, 450, 3008

\bibitem[\protect\citeauthoryear{Ercolano, Barlow, Storey  \& Liu}{Ercolano
  et~al.}{2003}]{Ercolano2003}
Ercolano B.,  Barlow M.~J.,  Storey P.~J.,   Liu X.~W.,  2003, \mn@doi [Monthly
  Notices of the Royal Astronomical Society]
  {10.1046/j.1365-8711.2003.06371.x}, 340, 1136

\bibitem[\protect\citeauthoryear{Ercolano, Barlow  \& Storey}{Ercolano
  et~al.}{2005}]{Ercolano2005a}
Ercolano B.,  Barlow M.~J.,   Storey P.~J.,  2005, \mn@doi [Monthly Notices of
  the Royal Astronomical Society] {10.1111/j.1365-2966.2005.09381.x}, 362, 1038

\bibitem[\protect\citeauthoryear{Ercolano, Young, Drake  \& Raymond}{Ercolano
  et~al.}{2008a}]{Ercolano2008}
Ercolano B.,  Young P.~R.,  Drake J.~J.,   Raymond J.~C.,  2008a, \mn@doi [The
  Astrophysical Journal Supplement Series] {10.1086/524378}, 175, 534

\bibitem[\protect\citeauthoryear{Ercolano, Drake, Raymond  \& Clarke}{Ercolano
  et~al.}{2008b}]{Ercolano2008a}
Ercolano B.,  Drake J.~J.,  Raymond J.~C.,   Clarke C.~C.,  2008b, \mn@doi [The
  Astrophysical Journal] {10.1086/590490}, 688, 398

\bibitem[\protect\citeauthoryear{Ercolano, Clarke  \& Drake}{Ercolano
  et~al.}{2009}]{Ercolano2009}
Ercolano B.,  Clarke C.~J.,   Drake J.~J.,  2009, \mn@doi [Astrophysical
  Journal] {10.1088/0004-637X/699/2/1639}, 699, 1639

\bibitem[\protect\citeauthoryear{Ercolano, Koepferl, Owen  \&
  Robitaille}{Ercolano et~al.}{2015}]{Ercolano2015}
Ercolano B.,  Koepferl C.,  Owen J.,   Robitaille T.,  2015, \mn@doi [Monthly
  Notices of the Royal Astronomical Society] {10.1093/mnras/stv1528}, 452, 3689

\bibitem[\protect\citeauthoryear{Ercolano, Picogna, Monsch, Drake  \&
  Preibisch}{Ercolano et~al.}{2021}]{Ercolano2021}
Ercolano B.,  Picogna G.,  Monsch K.,  Drake J.~J.,   Preibisch T.,  2021,
  \mn@doi [Monthly Notices of the Royal Astronomical Society]
  {10.1093/mnras/stab2590}, 508, 1675

\bibitem[\protect\citeauthoryear{Fang et~al.,}{Fang et~al.}{2018}]{Fang2018}
Fang M.,  et~al., 2018, \mn@doi [The Astrophysical Journal]
  {10.3847/1538-4357/aae780}, 868, 28

\bibitem[\protect\citeauthoryear{Fedele, Van Den~Ancker, Henning, Jayawardhana
  \& Oliveira}{Fedele et~al.}{2010}]{Fedele2010}
Fedele D.,  Van Den~Ancker M.~E.,  Henning T.,  Jayawardhana R.,   Oliveira
  J.~M.,  2010, \mn@doi [Astronomy and Astrophysics]
  {10.1051/0004-6361/200912810}, 510

\bibitem[\protect\citeauthoryear{Font, McCarthy, Johnstone  \& Ballantyne}{Font
  et~al.}{2004}]{Font2004}
Font A.~S.,  McCarthy I.~G.,  Johnstone D.,   Ballantyne D.~R.,  2004, \mn@doi
  [The Astrophysical Journal] {10.1086/383518}, 607, 890

\bibitem[\protect\citeauthoryear{Gangi et~al.,}{Gangi et~al.}{2020}]{Gangi2020}
Gangi M.,  et~al., 2020, \mn@doi [Astronomy and Astrophysics]
  {10.1051/0004-6361/202038534}, 643

\bibitem[\protect\citeauthoryear{Gressel, Turner, Nelson  \& McNally}{Gressel
  et~al.}{2015}]{Gressel2015a}
Gressel O.,  Turner N.~J.,  Nelson R.~P.,   McNally C.~P.,  2015, \mn@doi
  [Astrophysical Journal] {10.1088/0004-637X/801/2/84}, 801, 84

\bibitem[\protect\citeauthoryear{Gressel, Ramsey, Brinch, Nelson, Turner  \&
  Bruderer}{Gressel et~al.}{2020}]{Gressel2020a}
Gressel O.,  Ramsey J.~P.,  Brinch C.,  Nelson R.~P.,  Turner N.~J.,   Bruderer
  S.,  2020, \mn@doi [The Astrophysical Journal] {10.3847/1538-4357/ab91b7},
  896, 126

\bibitem[\protect\citeauthoryear{Haisch, Lada  \& Lada}{Haisch
  et~al.}{2001}]{Haisch2001}
Haisch Jr. K.~E.,  Lada E.~A.,   Lada C.~J.,  2001, \mn@doi [The Astrophysical
  Journal] {10.1086/320685}, 553, L153

\bibitem[\protect\citeauthoryear{Hamann}{Hamann}{1994}]{Hamann1994}
Hamann F.,  1994, \mn@doi [The Astrophysical Journal Supplement Series]
  {10.1086/192064}, 93, 485

\bibitem[\protect\citeauthoryear{Hartigan, Edwards  \& Ghandour}{Hartigan
  et~al.}{1995}]{Hartigan1995}
Hartigan P.,  Edwards S.,   Ghandour L.,  1995, \mn@doi [The Astrophysical
  Journal] {10.1086/176344}, 452, 736

\bibitem[\protect\citeauthoryear{Jennings, Ercolano  \& Rosotti}{Jennings
  et~al.}{2018}]{Jennings2018b}
Jennings J.,  Ercolano B.,   Rosotti G.~P.,  2018, \mn@doi [Monthly Notices of
  the Royal Astronomical Society] {10.1093/MNRAS/STY964}, 477, 4131

\bibitem[\protect\citeauthoryear{Kanagawa, Muto, Tanaka, Tanigawa, Takeuchi,
  Tsukagoshi  \& Momose}{Kanagawa et~al.}{2016}]{Kanagawa2016}
Kanagawa K.~D.,  Muto T.,  Tanaka H.,  Tanigawa T.,  Takeuchi T.,  Tsukagoshi
  T.,   Momose M.,  2016, \mn@doi [Publications of the Astronomical Society of
  Japan] {10.1093/pasj/psw037}, 68, 1

\bibitem[\protect\citeauthoryear{Klahr \& Kley}{Klahr \&
  Kley}{2006}]{Klahr2006}
Klahr H.,  Kley W.,  2006, \mn@doi [Astronomy and Astrophysics]
  {10.1051/0004-6361:20053238}, 445, 747

\bibitem[\protect\citeauthoryear{Kley, Bitsch  \& Klahr}{Kley
  et~al.}{2009}]{Kley2009}
Kley W.,  Bitsch B.,   Klahr H.,  2009, \mn@doi [Astronomy and Astrophysics]
  {10.1051/0004-6361/200912072}, 506, 971

\bibitem[\protect\citeauthoryear{Koepferl, Ercolano, Dale, Teixeira, Ratzka  \&
  Spezzi}{Koepferl et~al.}{2013}]{Koepferl2013}
Koepferl C.~M.,  Ercolano B.,  Dale J.,  Teixeira P.~S.,  Ratzka T.,   Spezzi
  L.,  2013, \mn@doi [Monthly Notices of the Royal Astronomical Society]
  {10.1093/mnras/sts276}, 428, 3327

\bibitem[\protect\citeauthoryear{Kudo, Hashimoto, Muto, Baobab~Liu, Dong,
  Hasegawa, Tsukagoshi  \& Konishi}{Kudo et~al.}{2018}]{Kudo2018}
Kudo T.,  Hashimoto J.,  Muto T.,  Baobab~Liu H.,  Dong R.,  Hasegawa Y.,
  Tsukagoshi T.,   Konishi M.,  2018, \mn@doi [The Astrophysical Journal]
  {10.3847/2041-8213/aaeb1c}, 868, L5

\bibitem[\protect\citeauthoryear{Liffman}{Liffman}{2003}]{Liffman2003}
Liffman K.,  2003, \mn@doi [Publications of the Astronomical Society of
  Australia] {10.1071/AS03019}, 20, 337

\bibitem[\protect\citeauthoryear{Manara et~al.,}{Manara
  et~al.}{2021}]{Manara2021}
Manara C.~F.,  et~al., 2021, \mn@doi [Astronomy and Astrophysics]
  {10.1051/0004-6361/202140639}, 650, A196

\bibitem[\protect\citeauthoryear{McGinnis, Dougados, Alencar, Bouvier  \&
  Cabrit}{McGinnis et~al.}{2018}]{McGinnis2018}
McGinnis P.,  Dougados C.,  Alencar S.~H.,  Bouvier J.,   Cabrit S.,  2018,
  \mn@doi [Astronomy and Astrophysics] {10.1051/0004-6361/201731629}, 620, A87

\bibitem[\protect\citeauthoryear{Mesa et~al.,}{Mesa et~al.}{2022}]{Mesa2022}
Mesa D.,  et~al., 2022, \mn@doi [Astronomy and Astrophysics]
  {10.1051/0004-6361/202142219}, 658

\bibitem[\protect\citeauthoryear{Mignone, Bodo, Massaglia, Matsakos, Tesileanu,
  Zanni  \& Ferrari}{Mignone et~al.}{2007}]{Mignone2007}
Mignone A.,  Bodo G.,  Massaglia S.,  Matsakos T.,  Tesileanu O.,  Zanni C.,
  Ferrari A.,  2007, \mn@doi [The Astrophysical Journal Supplement Series]
  {10.1086/513316}, 170, 228

\bibitem[\protect\citeauthoryear{Monsch, Ercolano, Picogna, Preibisch  \&
  Rau}{Monsch et~al.}{2019}]{Monsch2019}
Monsch K.,  Ercolano B.,  Picogna G.,  Preibisch T.,   Rau M.~M.,  2019,
  \mn@doi [Monthly Notices of the Royal Astronomical Society]
  {10.1093/mnras/sty3346}, 483, 3448

\bibitem[\protect\citeauthoryear{Monsch, Picogna, Ercolano  \&
  Preibisch}{Monsch et~al.}{2021}]{Monsch2021b}
Monsch K.,  Picogna G.,  Ercolano B.,   Preibisch T.,  2021, \mn@doi [Astronomy
  and Astrophysics] {10.1051/0004-6361/202140647}, 650, A199

\bibitem[\protect\citeauthoryear{Ogilvie \& Lubow}{Ogilvie \&
  Lubow}{2002}]{Ogilvie2002}
Ogilvie G.~I.,  Lubow S.~H.,  2002, \mn@doi [Monthly Notices of the Royal
  Astronomical Society] {10.1046/j.1365-8711.2002.05148.x}, 330, 950

\bibitem[\protect\citeauthoryear{Owen, Ercolano, Clarke  \& Alexander}{Owen
  et~al.}{2010}]{Owen2010}
Owen J.~E.,  Ercolano B.,  Clarke C.~J.,   Alexander R.~D.,  2010, \mn@doi
  [Monthly Notices of the Royal Astronomical Society]
  {10.1111/j.1365-2966.2009.15771.x}, 401, 1415

\bibitem[\protect\citeauthoryear{Owen, Clarke  \& Ercolano}{Owen
  et~al.}{2012}]{Owen2012}
Owen J.~E.,  Clarke C.~J.,   Ercolano B.,  2012, \mn@doi [Monthly Notices of
  the Royal Astronomical Society] {10.1111/j.1365-2966.2011.20337.x}, 422, 1880

\bibitem[\protect\citeauthoryear{Papaloizou \& Lin}{Papaloizou \&
  Lin}{1984}]{Papaloizou1984}
Papaloizou J.,  Lin D. N.~C.,  1984, \mn@doi [The Astrophysical Journal]
  {10.1086/162561}, 285, 818

\bibitem[\protect\citeauthoryear{Pascucci et~al.,}{Pascucci
  et~al.}{2020}]{Pascucci2020}
Pascucci I.,  et~al., 2020, \mn@doi [The Astrophysical Journal]
  {10.3847/1538-4357/abba3c}, 903, 78

\bibitem[\protect\citeauthoryear{Pascucci, Cabrit, Edwards, Gorti, Gressel  \&
  Suzuki}{Pascucci et~al.}{2022}]{Pascucci2022}
Pascucci I.,  Cabrit S.,  Edwards S.,  Gorti U.,  Gressel O.,   Suzuki T.,
  2022, arXiv e-prints, p. arXiv:2203.10068

\bibitem[\protect\citeauthoryear{Picogna, Ercolano, Owen  \& Weber}{Picogna
  et~al.}{2019}]{Picogna2019}
Picogna G.,  Ercolano B.,  Owen J.~E.,   Weber M.~L.,  2019, \mn@doi [Monthly
  Notices of the Royal Astronomical Society] {10.1093/mnras/stz1166}, 487, 691

\bibitem[\protect\citeauthoryear{Picogna, Ercolano  \& Espaillat}{Picogna
  et~al.}{2021}]{Picogna2021}
Picogna G.,  Ercolano B.,   Espaillat C.~C.,  2021, \mn@doi [Monthly Notices of
  the Royal Astronomical Society] {10.1093/mnras/stab2883}, 508, 3611

\bibitem[\protect\citeauthoryear{Pinte, Teague, Flaherty, Hall, Facchini  \&
  Casassus}{Pinte et~al.}{2022}]{Pinte2022}
Pinte C.,  Teague R.,  Flaherty K.,  Hall C.,  Facchini S.,   Casassus S.,
  2022, \mn@doi [arXiv e-prints] {10.48550/arxiv.2203.09528}, p.
  arXiv:2203.09528

\bibitem[\protect\citeauthoryear{Pontoppidan, Blake  \& Smette}{Pontoppidan
  et~al.}{2011}]{Pontoppidan2011}
Pontoppidan K.~M.,  Blake G.~A.,   Smette A.,  2011, \mn@doi [The Astrophysical
  Journal, Volume 733, Issue 2, article id. 84, <NUMPAGES>17</NUMPAGES> pp.
  (2011).] {10.1088/0004-637X/733/2/84}, 733, 84

\bibitem[\protect\citeauthoryear{Ribas, Mer{\'{i}}n, Bouy  \& Maud}{Ribas
  et~al.}{2014}]{Ribas2014}
Ribas {\'{A}}.,  Mer{\'{i}}n B.,  Bouy H.,   Maud L.~T.,  2014, \mn@doi
  [Astronomy and Astrophysics] {10.1051/0004-6361/201322597}, 561

\bibitem[\protect\citeauthoryear{Rigliaco, Pascucci, Gorti, Edwards  \&
  Hollenbach}{Rigliaco et~al.}{2013}]{Rigliaco2013}
Rigliaco E.,  Pascucci I.,  Gorti U.,  Edwards S.,   Hollenbach D.,  2013,
  \mn@doi [Astrophysical Journal] {10.1088/0004-637X/772/1/60}, 772

\bibitem[\protect\citeauthoryear{Rosotti, Ercolano, Owen  \& Armitage}{Rosotti
  et~al.}{2013}]{Rosotti2013}
Rosotti G.~P.,  Ercolano B.,  Owen J.~E.,   Armitage P.~J.,  2013, \mn@doi
  [Monthly Notices of the Royal Astronomical Society] {10.1093/mnras/sts725},
  430, 1392

\bibitem[\protect\citeauthoryear{Savage \& Sembach}{Savage \&
  Sembach}{1996}]{Savage1996}
Savage B.~D.,  Sembach K.~R.,  1996, \mn@doi [The Astrophysical Journal]
  {10.1086/177919}, 470, 893

\bibitem[\protect\citeauthoryear{Simon, Pascucci, Edwards, Feng, Gorti,
  Hollenbach, Rigliaco  \& Keane}{Simon et~al.}{2016}]{Simon2016a}
Simon M.~N.,  Pascucci I.,  Edwards S.,  Feng W.,  Gorti U.,  Hollenbach D.,
  Rigliaco E.,   Keane J.~T.,  2016, \mn@doi [The Astrophysical Journal]
  {10.3847/0004-637x/831/2/169}, 831, 169

\bibitem[\protect\citeauthoryear{Simon, Bai, Flaherty  \& Hughes}{Simon
  et~al.}{2018}]{Simon2018}
Simon J.~B.,  Bai X.-N.,  Flaherty K.~M.,   Hughes A.~M.,  2018, \mn@doi [The
  Astrophysical Journal] {10.3847/1538-4357/aad86d}, 865, 10

\bibitem[\protect\citeauthoryear{Wang, Bai  \& Goodman}{Wang
  et~al.}{2019}]{Wang2019}
Wang L.,  Bai X.-N.,   Goodman J.,  2019, \mn@doi [The Astrophysical Journal]
  {10.3847/1538-4357/ab06fd}, 874, 90

\bibitem[\protect\citeauthoryear{Weber, Ercolano, Picogna, Hartmann  \&
  Rodenkirch}{Weber et~al.}{2020}]{Weber2020}
Weber M.~L.,  Ercolano B.,  Picogna G.,  Hartmann L.,   Rodenkirch P.~J.,
  2020, \mn@doi [Monthly Notices of the Royal Astronomical Society]
  {10.1093/mnras/staa1549}, 496, 223

\makeatother
\end{thebibliography}



\appendix
\section{Line profiles and fits} \label{sec:appendix:profiles}

\begin{figure*}
    \centering
    \includegraphics[width=.8\textwidth]{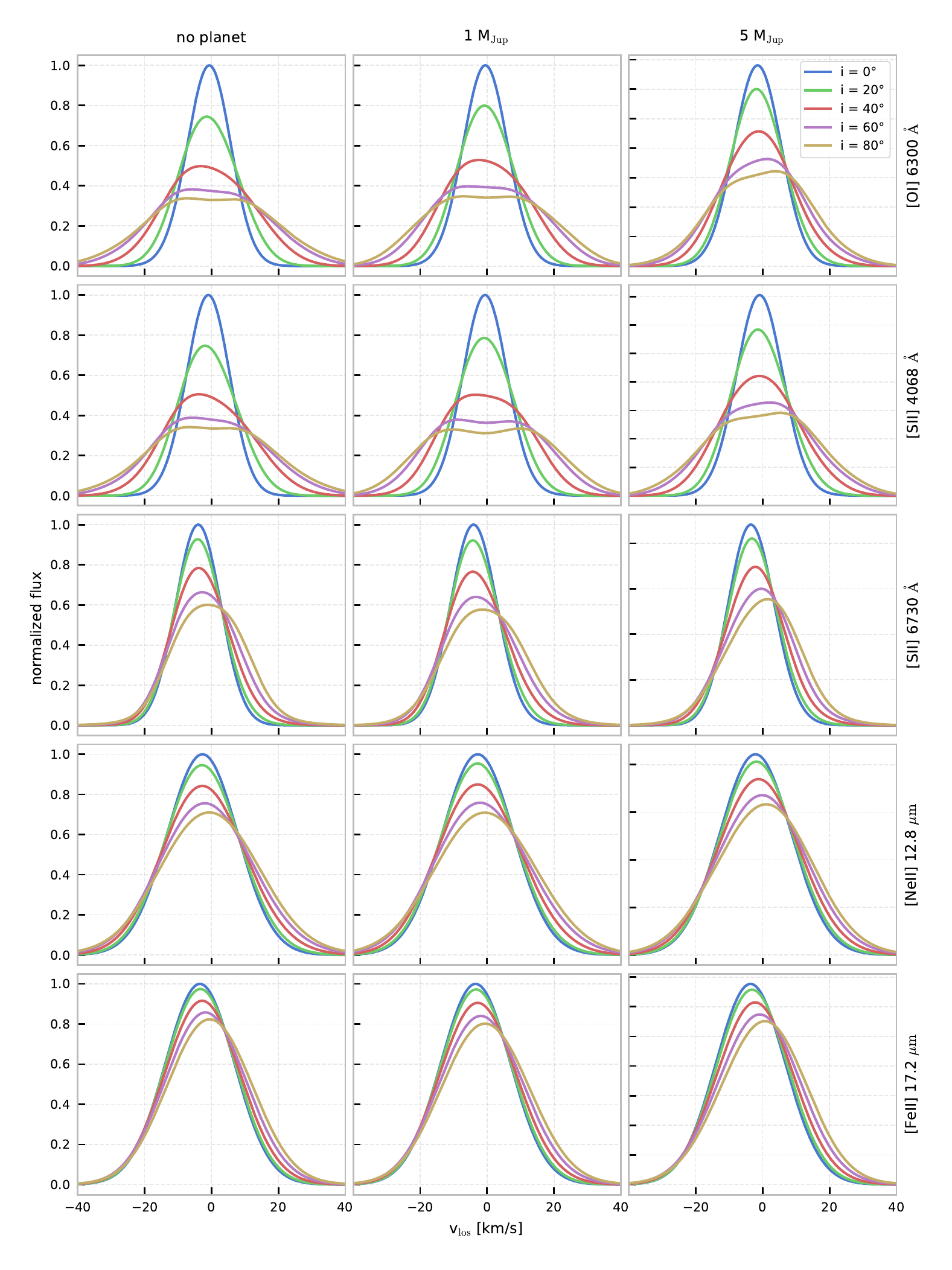}
    \caption{
        Synthetic spectral profiles viewed at inclination i and azimuth angle $\phi = 0$. The profiles are normalised to the peak of the face-on profile with i = 0\degree. The profiles were artificially degraded to a resolving power R = 50000 and R = 30000 for the optical and mid-infrared lines, respectively. This is comparable to currently available high-resolution observations.
    }
    \label{fig:profiles-all-current-res}
\end{figure*}

\begin{table*}
    \centering
    \caption{
        Results of the Gaussian decomposition of the [OI]~6300~\angstrom high-resolution (R=100000) line profiles. We also list the velocity of the actual peak of the profile v$_p$ in addition to the centroid velocity v$_c$. The velocities and FWHM are given in units of km/s. The results of all lines are available as online supplementary material.
    }
    \label{tab:fits_oi-6300}
    \begin{tabular}{ll|llll|llll|llll}
      &     & \multicolumn{4}{c}{MJ1} & \multicolumn{4}{c}{MJ5} & \multicolumn{4}{c}{REF} \\
    i[\degree] & phi [\degree] & v$_c$ & v$_p$ & FWHM & log(L/L$_\odot$) & v$_c$ & v$_p$ & FWHM & log(L/L$_\odot$) & v$_c$ & v$_p$ & FWHM & log(L/L$_\odot$)  \\
    \hline
    \hline
    0 & 0 & -0.50 & -0.38 & 9.82 & -5.04  & -1.50 & -1.38 & 11.16 & -4.88  & -0.62 & -0.38 & 9.42 & -4.94  \\
    \hline
    20 & 0 & -0.66 & -1.63 & 16.14 & -5.03  & -1.81 & -2.13 & 14.09 & -4.88  & -1.15 & -2.88 & 17.19 & -4.94  \\
     & 25 & -0.71 & -1.88 & 16.30 & -5.03  & -1.66 & -1.88 & 14.25 & -4.88  & & & &  \\
     & 50 & -0.70 & -2.13 & 16.80 & -5.03  & -1.55 & -1.63 & 14.71 & -4.88  & & & &  \\
     & 75 & -0.64 & -1.88 & 17.54 & -5.03  & -1.54 & -1.63 & 15.42 & -4.88  & & & &  \\
     & 100 & -0.61 & -1.88 & 18.29 & -5.04  & -1.58 & -2.13 & 16.02 & -4.88  & & & &  \\
     & 125 & -0.64 & -2.38 & 18.81 & -5.03  & -1.62 & -2.38 & 16.40 & -4.88  & & & &  \\
     & 150 & -0.78 & -2.88 & 18.78 & -5.03  & -1.64 & -2.38 & 16.70 & -4.88  & & & &  \\
     & 175 & -1.01 & -2.88 & 18.07 & -5.03  & -1.68 & -2.38 & 16.77 & -4.88  & & & &  \\
     & 200 & -1.23 & -2.63 & 17.15 & -5.03  & -1.76 & -2.63 & 16.26 & -4.88  & & & &  \\
     & 225 & -1.31 & -2.63 & 16.27 & -5.03  & -1.86 & -2.38 & 15.46 & -4.88  & & & &  \\
     & 250 & -1.19 & -2.13 & 15.67 & -5.03  & -1.94 & -2.13 & 14.90 & -4.88  & & & &  \\
     & 275 & -0.96 & -1.38 & 15.63 & -5.03  & -2.00 & -2.13 & 14.52 & -4.88  & & & &  \\
     & 300 & -0.75 & -1.38 & 15.81 & -5.03  & -2.04 & -2.38 & 14.26 & -4.88  & & & &  \\
     & 325 & -0.64 & -1.38 & 16.03 & -5.03  & -2.00 & -2.38 & 14.11 & -4.88  & & & &  \\
     & 350 & -0.64 & -1.63 & 16.13 & -5.03  & -1.87 & -2.13 & 14.07 & -4.88  & & & &  \\
    \hline
    40 & 0 & -0.66 & -7.14 & 27.84 & -5.03  & -1.45 & 1.38 & 21.77 & -4.88  & -1.16 & -7.14 & 28.53 & -4.94  \\
     & 25 & -0.66 & -6.39 & 28.13 & -5.03  & -1.25 & 0.63 & 21.66 & -4.88  & & & &  \\
     & 50 & -0.57 & 1.88 & 28.50 & -5.04  & -1.19 & 0.88 & 22.18 & -4.88  & & & &  \\
     & 75 & 4.91 & 4.39 & 19.81 & -5.21  & -1.29 & 1.63 & 23.72 & -4.88  & & & &  \\
     &  & -10.45 & 4.39 & 12.86 & -5.52  & & & &  & & & &  \\
     & 100 & 6.85 & -7.64 & 17.61 & -5.28  & -1.43 & -6.14 & 24.71 & -4.88  & & & &  \\
     &  & -9.36 & -7.64 & 14.68 & -5.39  & & & &  & & & &  \\
     & 125 & 8.33 & -7.39 & 16.83 & -5.35  & -1.48 & 1.63 & 25.35 & -4.88  & & & &  \\
     &  & -8.42 & -7.39 & 16.87 & -5.32  & & & &  & & & &  \\
     & 150 & 8.65 & -7.64 & 17.57 & -5.36  & -1.49 & 2.13 & 25.14 & -4.88  & & & &  \\
     &  & -8.21 & -7.64 & 17.69 & -5.30  & & & &  & & & &  \\
     & 175 & -0.97 & -6.89 & 29.88 & -5.03  & -1.55 & 1.88 & 25.22 & -4.88  & & & &  \\
     & 200 & -1.29 & -6.14 & 28.25 & -5.03  & -1.74 & -3.88 & 23.27 & -4.87  & & & &  \\
     & 225 & -1.43 & -5.39 & 27.11 & -5.03  & -1.87 & -2.63 & 21.71 & -4.88  & & & &  \\
     & 250 & -1.29 & -5.39 & 26.73 & -5.03  & -1.84 & -1.63 & 20.70 & -4.89  & & & &  \\
     & 275 & -1.00 & -6.39 & 26.98 & -5.03  & -1.81 & -1.13 & 20.61 & -4.88  & & & &  \\
     & 300 & -0.77 & -6.89 & 27.41 & -5.03  & -1.82 & -1.13 & 20.79 & -4.88  & & & &  \\
     & 325 & -0.66 & -7.14 & 27.77 & -5.03  & -1.72 & 0.38 & 21.21 & -4.88  & & & &  \\
     & 350 & -0.66 & -7.14 & 27.77 & -5.03  & -1.53 & 1.38 & 21.67 & -4.88  & & & &
\end{tabular}
\end{table*}

\begin{table*}
    \centering
    \contcaption{Results of the Gaussian decomposition of the [OI]~6300~\angstrom line profiles.}
    \begin{tabular}{ll|llll|llll|llll}
      &     & \multicolumn{4}{c}{MJ1} & \multicolumn{4}{c}{MJ5} & \multicolumn{4}{c}{REF} \\
    i [\degree] & phi [\degree] & v$_c$ & v$_p$ & FWHM & log(L/L$_\odot$) & v$_c$ & v$_p$ & FWHM & log(L/L$_\odot$) & v$_c$ & v$_p$ & FWHM & log(L/L$_\odot$)  \\
    \hline
    \hline
    60 & 0 & 8.79 & -9.40 & 23.55 & -5.27  & -0.77 & 4.89 & 28.08 & -4.87  & -0.76 & -9.65 & 37.20 & -4.94  \\
     &  & -11.60 & -9.40 & 18.33 & -5.41  & & & &  & & & &  \\
     & 25 & 8.37 & -9.15 & 24.33 & -5.26  & -0.53 & 3.63 & 27.75 & -4.87  & & & &  \\
     &  & -11.61 & -9.15 & 18.94 & -5.42  & & & &  & & & &  \\
     & 50 & -0.14 & 5.39 & 36.44 & -5.03  & -0.49 & 3.38 & 27.70 & -4.87  & & & &  \\
     & 75 & 7.81 & 7.39 & 23.14 & -5.24  & -0.70 & 4.89 & 29.40 & -4.88  & & & &  \\
     &  & -12.54 & 7.39 & 17.40 & -5.46  & & & &  & & & &  \\
     & 100 & 10.00 & 8.90 & 20.19 & -5.32  & -0.88 & 5.64 & 30.63 & -4.88  & & & &  \\
     &  & -11.09 & 8.90 & 19.10 & -5.36  & & & &  & & & &  \\
     & 125 & 12.42 & -9.65 & 18.24 & -5.41  & 5.20 & 5.14 & 20.51 & -5.07  & & & &  \\
     &  & 1.82 & -9.65 & 10.08 & -6.39  & -11.42 & 5.14 & 14.09 & -5.34  & & & &  \\
     &  & -10.39 & -9.65 & 21.10 & -5.31  & & & &  & & & &  \\
     & 150 & 11.00 & -10.65 & 21.95 & -5.34  & 5.29 & 5.64 & 20.14 & -5.08  & & & &  \\
     &  & -10.83 & -10.65 & 21.79 & -5.32  & -11.60 & 5.64 & 13.97 & -5.33  & & & &  \\
     & 175 & 9.89 & -9.15 & 24.06 & -5.31  & 5.90 & -7.64 & 19.99 & -5.11  & & & &  \\
     &  & -11.33 & -9.15 & 21.17 & -5.35  & -10.76 & -7.64 & 15.51 & -5.28  & & & &  \\
     & 200 & -0.99 & -8.15 & 36.85 & -5.02  & -1.35 & -4.64 & 28.98 & -4.87  & & & &  \\
     & 225 & -1.16 & -7.39 & 35.69 & -5.03  & -1.49 & -3.88 & 27.28 & -4.88  & & & &  \\
     & 250 & -1.10 & -8.15 & 34.81 & -5.02  & -1.36 & -1.38 & 26.48 & -4.88  & & & &  \\
     & 275 & 7.82 & -9.65 & 22.81 & -5.27  & -1.22 & 2.13 & 27.42 & -4.88  & & & &  \\
     &  & -11.58 & -9.65 & 17.37 & -5.42  & & & &  & & & &  \\
     & 300 & 8.18 & -10.65 & 22.61 & -5.27  & -1.19 & 3.38 & 28.26 & -4.88  & & & &  \\
     &  & -11.95 & -10.65 & 16.89 & -5.42  & & & &  & & & &  \\
     & 325 & 8.63 & -10.90 & 23.02 & -5.27  & -1.10 & 4.14 & 27.90 & -4.87  & & & &  \\
     &  & -11.92 & -10.90 & 17.63 & -5.41  & & & &  & & & &  \\
     & 350 & 8.94 & -9.40 & 23.12 & -5.28  & -0.88 & 4.89 & 28.12 & -4.87  & & & &  \\
     &  & -11.60 & -9.40 & 18.13 & -5.40  & & & &  & & & & \\
    \hline
    80 & 0 & 11.15 & -9.40 & 25.08 & -5.30  & -0.01 & 6.89 & 32.20 & -4.87  & 11.04 & -9.90 & 25.84 & -5.23  \\
     &  & -12.40 & -9.40 & 22.49 & -5.37  & & & &  & -12.11 & -9.90 & 23.72 & -5.28  \\
     & 25 & 10.88 & 7.89 & 25.43 & -5.30  & 0.27 & 4.89 & 31.38 & -4.87  & & & &  \\
     &  & -12.05 & 7.89 & 23.33 & -5.37  & & & &  & & & &  \\
     & 50 & 0.42 & 7.39 & 40.23 & -5.02  & 0.32 & 4.64 & 30.70 & -4.87  & & & &  \\
     & 75 & 10.09 & 9.15 & 23.89 & -5.28  & 0.06 & 6.14 & 31.77 & -4.88  & & & &  \\
     &  & -12.67 & 9.15 & 21.24 & -5.40  & & & &  & & & &  \\
     & 100 & 12.08 & 11.15 & 20.68 & -5.35  & -0.13 & 6.89 & 32.77 & -4.88  & & & &  \\
     &  & -11.27 & 11.15 & 22.26 & -5.33  & & & &  & & & &  \\
     & 125 & 13.47 & -10.15 & 20.53 & -5.39  & -0.19 & 7.14 & 33.06 & -4.87  & & & &  \\
     &  & -10.58 & -10.15 & 24.69 & -5.28  & & & &  & & & &  \\
     & 150 & 16.52 & -11.40 & 17.93 & -5.52  & 8.25 & 7.64 & 19.09 & -5.16  & & & &  \\
     &  & 5.23 & -11.40 & 12.75 & -5.98  & -10.21 & 7.64 & 17.10 & -5.23  & & & &  \\
     &  & -10.89 & -11.40 & 24.87 & -5.28  & & & &  & & & &  \\
     & 175 & 11.88 & 8.40 & 24.63 & -5.33  & 9.62 & -7.64 & 18.12 & -5.22  & & & &  \\
     &  & -12.04 & 8.40 & 24.00 & -5.33  & -8.85 & -7.64 & 18.81 & -5.16  & & & &  \\
     & 200 & 7.68 & -8.65 & 29.48 & -5.21  & -0.73 & -5.14 & 31.98 & -4.87  & & & &  \\
     &  & -9.54 & -8.65 & 11.65 & -5.89  & & & &  & & & &  \\
     &  & -19.97 & -8.65 & 17.27 & -5.74  & & & &  & & & &  \\
     & 225 & -0.70 & -7.89 & 40.25 & -5.02  & -0.86 & -4.64 & 30.58 & -4.88  & & & &  \\
     & 250 & -0.70 & -8.90 & 39.68 & -5.02  & -0.66 & -5.39 & 30.68 & -4.88  & & & &  \\
     & 275 & 10.45 & -10.90 & 23.39 & -5.31  & -0.46 & 5.39 & 31.78 & -4.88  & & & &  \\
     &  & -12.22 & -10.90 & 19.96 & -5.37  & & & &  & & & &  \\
     & 300 & 10.04 & -12.66 & 24.73 & -5.28  & -0.41 & -7.14 & 32.25 & -4.88  & & & &  \\
     &  & -13.35 & -12.66 & 19.18 & -5.41  & & & &  & & & &  \\
     & 325 & 10.54 & 10.65 & 25.41 & -5.28  & -0.33 & 6.89 & 32.34 & -4.87  & & & &  \\
     &  & -13.31 & 10.65 & 20.78 & -5.39  & & & &  & & & &  \\
     & 350 & 11.23 & -9.15 & 24.83 & -5.30  & 7.50 & 7.14 & 18.62 & -5.14  & & & &  \\
     &  & -12.52 & -9.15 & 22.12 & -5.37  & -10.51 & 7.14 & 16.56 & -5.26  & & & & 
\end{tabular}
\end{table*}

\label{sec:appendix}


\bsp	
\label{lastpage}
\end{document}